\documentclass[mnsc,nonblindrev]{informs3}

\OneAndAHalfSpacedXI

\usepackage[plainpages=false,hyperfootnotes=false]{hyperref}
\hypersetup{
  colorlinks   = true, 
  urlcolor     = blue, 
  linkcolor    = red, 
  citecolor   = blue 
}
\newcommand{\new}[1]{{\color{black} #1}}

\usepackage{natbib}
 \bibpunct[, ]{(}{)}{,}{a}{}{,}%
 \def\bibfont{\small}%
\usepackage{dsfont}
\usepackage[colorinlistoftodos]{todonotes}
\usepackage{enumitem}
\usepackage{booktabs}
\usepackage{multirow}
\usepackage{lscape}
\usepackage{wrapfig}
\usepackage{changepage}
 
\newcommand{\tU}{{\tilde{U}}}

\newcommand{\C}{{\mathbb{C}}}
\newcommand{\E}{{\mathbb{E}}}
\newcommand{\Q}{{\mathbb{Q}}}
\newcommand{\V}{{\mathbb{V}}}
\renewcommand{\P}{{\mathbb{P}}}
\newcommand{\Lp}{{\mathbb{L}^2}}

\newcommand{\Id}{{\mathds{1}}}
\newcommand{\R}{{\mathds{R}}}

\newcommand{\tT}{{t\in[0,T]}}

\newcommand{\A}{{\mathcal{A}}}

\newcommand{\Co}{{\textrm{C}_{\pi, \delta}}}
\newcommand{\CoIn}{{\textrm{C}_{\pi, \delta}}}

\newcommand{\Con}{{\textrm{C}_{\pi| \delta}}}
\newcommand{\Coninv}{ \;\breve{{\!\!\textrm{C}}}_{\pi| \delta} }
\renewcommand{\C}{{\mathcal{C}}}
\newcommand{\D}{{\mathcal{D}}}
\newcommand{\F}{{\mathcal{F}}}
\newcommand{\B}{{\mathcal{B}}}
\newcommand{\M}{{\mathcal{M}}}

\renewcommand{\L}{{\mathcal{L}}}
\newcommand{\U}{{\mathcal{U}}}

\newcommand{\m}{{\mathfrak{m}}}
\newcommand{\s}{{\mathfrak{s}}}

\newcommand{\mC}{{\mathfrak{C}}}
\newcommand{\RM}{{\mathcal{R}}}

\newcommand{\ep}{\varepsilon}
\newcommand{\ml}{\underline{m}}
\newcommand{\mr}{\overline{m}}
\newcommand{\sl}{\underline{s}}
\newcommand{\sr}{\overline{s}}

\newcommand{\TVaR}{{\text{TVaR}}}
\newcommand{\UTE}{{\text{UTE}}}

\newcommand{\pis}{{\pi^*}}
\newcommand{\pist}{{\pi^\star}}

\newcommand{\norm}[1]{\left\lVert#1\right\rVert}

\renewcommand{\Finv}{{\breve{F}}}

\newcommand{\oM}{{\breve{\M}}}

\TheoremsNumberedThrough     
\ECRepeatTheorems

\EquationsNumberedThrough    

\MANUSCRIPTNO{MS-0001-1922.65}

\begin{document}

\VOLUME{}%
\NO{}%
\MONTH{}
\YEAR{}
\FIRSTPAGE{}%
\LASTPAGE{}%
\SHORTYEAR{}
\ISSUE{} %
\LONGFIRSTPAGE{} %
\DOI{}%

\RUNAUTHOR{Pesenti S. M. and Jaimungal S.}

\RUNTITLE{Portfolio Optimisation within a Wasserstein Ball}

\TITLE{Portfolio Optimisation within a Wasserstein Ball}

\ARTICLEAUTHORS{%
\AUTHOR{Silvana M. Pesenti}
\AFF{Department of Statistical Sciences, University of Toronto \EMAIL{silvana.pesenti@utoronto.ca}} 
\AUTHOR{Sebastian Jaimungal}
\AFF{Department of Statistical Sciences, University of Toronto \EMAIL{sebastian.jaimungal@utoronto.ca}}
} 

\ABSTRACT{%
We study the problem of active portfolio management where an investor aims to outperform a benchmark strategy's risk profile while not deviating too far from it. Specifically, an investor considers alternative strategies whose terminal wealth lie within a Wasserstein ball surrounding a benchmark's -- being distributionally close -- and that have a specified dependence/copula -- tying state-by-state outcomes -- to it. The investor then chooses the alternative strategy that minimises a distortion risk measure of terminal wealth. In a general (complete) market model, we prove that an optimal dynamic strategy exists and provide its characterisation through the notion of isotonic projections. 

We further propose a simulation approach to calculate the optimal strategy's terminal wealth, making our approach applicable to a wide range of market models. Finally, we illustrate how investors with different copula and risk preferences invest and improve upon the benchmark using the Tail Value-at-Risk, inverse S-shaped, and lower- and upper-tail distortion risk measures as examples. We find that investors' optimal terminal wealth distribution has larger probability masses in regions that reduce their risk measure relative to the benchmark while preserving the benchmark's structure.
}%



\maketitle

%

\section{Introduction}

The existence of the plethora of exchange traded funds (ETFs) that track particular indices shows that investors are concerned about benchmarks. Institutional investors, also, often aim to track indices such as the Standard and Poors 500 index and the Down Jones Industrial Average. The terminal wealth of these strategies or benchmarks, however, may not perfectly align with the investors' risk preferences. To adjust the terminal wealth of these benchmark strategies to their risk preferences, investors may be willing to deviate from such benchmarks  -- but not too far. For example, investors may wish to better protect themselves from the benchmark's downside risk or seek larger gains which are not attainable using the benchmark. They do not, however, want to abandon the benchmark entirely and settle for a strategy that provides maximal downside risk protection or maximal upside gains. Instead, investors aim to stay close to a benchmark's terminal wealth, but allow for enough flexibility to (partially) adjust the benchmark's wealth to their personal risk preferences, captured via a distortion risk measure. Distortion risk measures include a wide range of risk measures used in behavioural economics to describe investors' risk preferences \citep{Tversky1992JRU} and the most widely used risk measure in financial risk management: the Tail Value-at-Risk (TVaR) \citep{Acerbi2002JBF}. 


A further crucial concern of investors is that their strategy may underperform the benchmark's terminal wealth. For an alternative strategy whose terminal wealth distribution is close to that of the benchmark's, investors can still face a situation where the benchmark and the alternative strategy differ considerably on any set of outcomes. For example, scenarios may occur where the benchmark provides positive returns while the alternatives' are negative. Thus, we allow the investor to specify their choice of dependence structure / copula with the benchmark \new{(a state-by-state preference)}, e.g., they may choose to co-move with the benchmark, be independent of the benchmark, co-move with the benchmark above a threshold but be independent below a threshold, and so on. 
A related set of works, in a different context, that incorporates a \new{prespecified} copula when seeking over cost-efficient portfolios for utility maximisers may be found in \cite{bernard2014explicit,bernard2014optimal,bernard2015optimal}.

To quantify the discrepancy between the distributions of the benchmark's and alternative strategies' terminal wealth, we utilise the Wasserstein distance \citep{Villani2008book}. The Wasserstein distance allows for comparisons between distributions with differing supports and has been widely applied in finance and economics to represent model uncertainty; we refer to the review by \cite{pflug2018review}. \new{In the present context, being able to measure distances between distributions with differing supports is important, as it is apriori not obvious whether the benchmark's and the optimal's terminal distribution share the same support.} Robustifying distortion risk measures within a Wasserstein-ball and additional mean and variance constraint is studied in \cite{Bernard2020}. In our context, investors seek alternative strategies whose terminal wealth distribution lies within an $\ep$-Wasserstein distance from the benchmark's, where the parameter $\ep$ reflects an investor's tolerance to diverge from the benchmark. By choosing a positive tolerance level $\ep$, investors can control the trade-off between staying close to the benchmark and minimising their distortion risk measure. When $\ep$ drops to zero, the investor obtains a distribution that matches the benchmark's, but may on a scenario by scenario basis differ from it -- contingent on their choice of copula. When $\ep$ tends to infinity, the investor seeks over all strategies and views them all as equally viable; this limit corresponds to the classical optimal portfolio selection problem, but with a particular choice of dependence structure.

In this paper, we aim to address the investors' considerations raised in the preceding paragraphs. Thus, we solve the optimal portfolio problem where investors seek strategies that (a) satisfy a budget constraint -- they are at most as expensive as the benchmark, (b) have terminal wealth within an $\ep$-Wasserstein ball around the benchmark -- they are distributionally close, (c) have terminal wealth that has a specific dependence structure with the benchmark's -- they have a specified copula, and (d) improve upon the benchmark's distortion risk measure -- they have better risk-reward profiles. 


The main contribution of this paper is the solution to the investor's portfolio choice problem, which we tackle by recasting it as a constrained convex optimisation over quantile functions. We prove that the quantile formulation, and hence (under additional weak assumptions) the original optimisation problem, has a unique solution. We then characterise the quantile function of the optimal terminal wealth via an isotonic projection (see e.g., \cite{nemeth2016isotonic}) of a linear combination of the benchmark's terminal wealth \new{quantile function}, the \new{weight function of the} distortion risk measure, and a conditional stochastic discount factor (SDF). We further show that the investor's optimal strategy derives from replicating a random payoff that depends solely on the benchmark and the SDF. We discuss investors' choices of the $\ep$-Wasserstein ball and provide a simulation approach for obtaining the optimal terminal wealth distribution for any (complete) market model. Furthermore, we solve the optimisation problem when the investor does not specify a copula. Through numerical experiments, we illustrate how investors with different risk measures and copula choices optimally invest.

As the investor aims to outperform a benchmark (while simultaneously remaining close to it), our work pertains to active portfolio management. This contrasts with passive portfolio management, which involves tracking a given benchmark strategy, and with optimal portfolio allocation, where benchmarks are absent.
Active portfolio management was first formalised in \cite{browne1999beating}, \cite{browne1999reaching}, and \cite{browne2000risk}, where the authors study an investor who trades in a GBM market model and aims to maximise the relative performance with respect to a stochastic benchmark. \cite{basak2006risk} studies portfolio allocation problem where the investor with constant relative risk aversion (CRRA) aims to maximise expected utility of terminal wealth in a GBM market model, subject to not underperforming a benchmark's return. 
\cite{grith2017reference} propose benchmarking to partially explain the empirical pricing kernel puzzle by assuming that investors switch utilities depending on their relative performance to a benchmark. 
\cite{al2018outperformance,al2021active} solve for the optimal portfolio allocation, where the agent aims to track closely, but outperform in log-utility,  a benchmark. For a comprehensive overview of active portfolio management see \cite{grinold2019advances}. 
Our work differs from these studies in that we seek alternative strategies that  optimise an investor's distortion risk measure of their terminal wealth, while remaining close to the benchmark in the Wasserstein distance and via a chosen copula. Furthermore, the approach we develop is applicable to any (complete) market model that the investor can simulate.

Representing investors  risk attitudes via distortion risk measures has been considered in portfolio selection since \cite{Rockafellar2000JR}; however, non-convex risk measures such as inverse S-shaped (IS) probability distortions have only recently been studied. \cite{Jin2008MF}, for example, consider a continuous-time portfolio selection problem under cumulative prospect theory (CPT) \citep{Tversky1992JRU}; these authors use IS risk measures together with an utility function, and show that the resulting optimum is attained by means of terminal payoffs of the binary type. \cite{He2011MS} consider static optimisation problems where an investor decides on how to invest in a single risky asset to optimise their CPT risk measure. Our portfolio optimisation problem, while conceptually different, also applies to CPT with linear utility, i.e., IS risk measures. 


Applications of the Wasserstein distance in portfolio optimisation focus on model or distributional uncertainty, rather than on deviations from a benchmark. Specifically, distributionally robust portfolio optimisation concerns portfolio optimisation when the underlying distribution is only partially specified; for example, see \cite{pflug2007ambiguity, Wozabal2014OR, gao2016distributionally, pflug2018review, esfahani2018data, Blanchet2018SSRN, ji2018data}, and \cite{Blanchet2019MOR}. These articles, while utilising the Wasserstein distance for portfolio selection, do not consider a benchmark strategy, and the Wasserstein distance pertains to model uncertainty  constructed in a data-driven way via, e.g., the empirical distribution.

When the Wasserstein constraint is absent ($\ep \to\infty$ in our formulation), the benchmark's distribution becomes redundant. If the copula constraint \new{in our optimisation problem} is also lifted, then our problem formulation reduces to finding a strategy that minimises a risk measure subject to a budget constraint. \cite{Schied2004AAP} study this special case with almost surely bounded terminal wealth, and \cite{he2011portfolio} when the terminal wealth is bounded from below, and show that for distortion risk measures, the optimal terminal wealth distribution consists of two point masses. \new{Our approach to solving our portfolio optimisation problem builds on the
\cite{he2011portfolio}'s reformulation of their portfolio optimisation via an optimisation over quantile functions. In this manuscript, however, we additionally consider a prespecified copula and a Wasserstein constraint. 
}
\cite{ruschendorf2019construction} extend the results of \cite{he2011portfolio} and show that when the terminal wealth's support is compact, the optimal distribution has support on three points. Neither of these optimal distributions is satisfying from a practical perspective; by contrast, while our optimal distributions may have probability masses, it does not entirely consist of them and it partially inherits the benchmark's terminal distribution structure.


The remainder of the paper is organised as follows: In Section \ref{sec: problem-formulation} we introduce the investor's optimisation problem. To solve this non-convex optimisation problem, we first transform the optimisation into one over a quantile function, as described in Section \ref{subsec: opt quant}. In Section \ref{sec: all-coupla} we consider the optimisation problem when an investor specifies a copula between alternative strategies and the benchmark. We discuss the choice of the Wasserstein distance in Section \ref{sec: choice-of-ep} and provide a simulation approach (Section \ref{sec:simualtion-approach}) to calculate the optimal terminal wealth from simulations of a market model. The optimal terminal wealth random variables for investors with different risk measures and choices of copulas is illustrated in Section \ref{sec: simulation-example}.

\section{Problem Formulation}\label{sec: problem-formulation}

In this section, we introduce the market model and develop  the investor's optimisation problem.

\subsection{The Market Model}

We work on a (completed) filtered probability space $(\Omega,\P,\F=\{\F_t\}_{\tT},\mathbb{F})$, where $\F$ is the natural filtration generated by the underlying processes. We denote the set of risky asset price processes by $S=(S_t^1,\dots,S_t^d)_{t\ge0}$ and denote the stochastic discount factor (SDF)  by $\varsigma=(\varsigma_t)_{t\ge0}$. Thus, $(\varsigma_t\,S_t^i)_{t\ge0}$ is a $\P$-martingale, for all $i\in\D:=\{1,\dots,d\}$. We assume the market is complete and for simplicity that the SDF has continuous distribution function -- this assumption can  be removed at the cost of some added complexities in the construction of our optimal strategy.

An investor chooses from the set of admissible strategies $\A$ consisting of $\F$-predictable Markov processes that are self-financing and in $\Lp(\Omega,[0,T])$. For an arbitrary $\pi = (\pi_t^1, \ldots, \pi_t^d)_{\tT}\in\A$, the investor's wealth process  $X^\pi=(X^\pi_t)_\tT$ satisfies the stochastic differential equation (SDE)
\begin{equation}
dX_t^\pi = X_t^\pi (1-\mathbf{1}^\intercal\,\pi_t)\,r_t\,dt + X_t^\pi
\sum_{i\in\D} \pi_t^i\, \frac{dS_t^i}{S_t^i}\,;
\label{eqn:dX_upsilon}
\end{equation}
hence, $\pi$ denotes the proportion of wealth invested in the risky assets.

\noindent
\begin{example}[SIR-CEV Market Model]
\label{ex: SIR-MM}
\begin{small}
To illustrate the flexibility of our formulation, we use as a running example a stochastic interest rate model combined with a constant elasticity of variance (SIR-CEV) market model. For this market model, the filtration is generated by a $d$-dimensional $\P$-Brownian motion $W=(W_t)_\tT$ with correlation structure $d[W^i,W^j]_t=\rho_{ij}\,dt$, for $i,j\in\D$, and we use $\rho$ to denote the matrix whose entries are $\rho_{ij}$. The first $(d-1)$ risky assets are equities and satisfy the SDEs
\begin{subequations}
\begin{equation}
    dS_t^i = S_t^i\left(\mu^i\,dt + \sigma^i\, (S_t^i)^{\beta^i}\,dW_t^i \right), \qquad  i\in\D/\{d\},
    \label{eqn:dS-GBM}
\end{equation}
where $\mu=(\mu^1,\dots,\mu^{d-1})$ is a vector of drifts,  $\sigma=(\sigma^1,\dots,\sigma^{d-1})$ is a vector of (strictly positive) instantaneous volatilities, and $\beta=(\beta^1,\dots,\beta^{d-1})$ are the CEV parameters. The $d^{th}$ risky asset is the $T$-bond, and satisfies the SDE
\begin{equation}
dS_t^d = S_t^d\left( \left((1+b\,B_t) r_t - a\,B_t \right)\,dt - \sigma_r\,B_t\,dW_t^d\right),
\end{equation}%
where $B_t=(1-e^{-\kappa(T-t)})/\kappa$, $a=\kappa\theta-\hat{\kappa}\hat{\theta}$, $b=\kappa-\hat{\kappa}$, and the stochastic interest rate satisfies the SDE
\begin{equation}
    dr_t = \kappa(\theta-r_t)\,dt+\sigma_r\,dW_t^d.
\end{equation}%
\end{subequations}%
Finally, the SDF $\varsigma$ satisfies the SDE
\begin{equation}
    d\varsigma_t = -\varsigma_t\,\left(r_t\,dt+ \lambda_t^\intercal\,dW_t\right)\,,
    \label{eqn:sdf-gbm}
\end{equation}
where $\lambda=(\lambda_t)_\tT$ denotes the vector-valued market price of risk with  $(\rho \lambda_t)^i =\frac{\mu^i-r_t}{\sigma^i(S_t^i)}$ for $i\in\D/\{d\}$, and
$(\rho \lambda_t)^d =\frac{a-b\,r_t}{\sigma_r}$. By construction, the process $(\varsigma_tS_t^i)_{\tT}$ is a $\P$-martingale for all $i\in\D$.
\end{small}
\end{example}
\vspace{1em} 

\subsection{The Benchmark Strategy}
\label{sec: The benchmark strategy}

An investor chooses a benchmark strategy $\delta\in\A$ which they aim to improve upon. The benchmark's terminal wealth  $X_T^\delta$ is assumed to be square-integrable with a continuous distribution function. We call $F$ its $\P$-distribution function and write $X_T^\delta \stackrel{\P}{\sim}F$. Furthermore, we define its quantile function (left-inverse) as $\Finv(u) := \inf\{y \in \R ~|~ F(y) \ge u\}$, $u \in [0,1] $, where we set $\inf\emptyset := + \infty$. The mean and standard deviation of the benchmark strategy are denoted
\begin{equation}
    \m := \E[X_T^\delta] 
    \qquad \text{and} \qquad
    \s := \sqrt{\V[X_T^\delta]\;}.
\end{equation}
Typical examples of benchmark strategies include (i) constant proportion strategies, such as $30\%$ in foreign risky assets, $30\%$ in domestic risk assets, and $40\%$ in the bank account; (ii) deterministic proportion strategies, such as one in which the proportions start as in (i) but annually ratchet towards  $0\%$ in foreign risky assets, $10\%$ in domestic risk assets, and $90\%$ in the bank account; and (iii) market (capitalisation weighted) portfolios.

Given a benchmark strategy $\delta\in\A$, the investor considers strategies $\pi\in\A$ whose terminal wealth $X_T^\pi$ and that of the benchmark's $X_T^\delta$ move according to a specified dependence, such that e.g., if the benchmark performs well so do the alternatives. More accurately, the investor \new{specifies their state-by-state preference and }considers only strategies such that $(X_T^\pi, X_T^\delta)$ have a copula $\Co$. Additionally, the investor requires the terminal wealth distribution of each considered alternative to be close to the benchmark's terminal wealth's distribution, where closeness is defined in the sense of the Wasserstein distance. To formalise these notions, we recall the notion of copula and the Wasserstein distance below. 

\begin{definition}[\cite{Sklar1959SP}]
A random vector $(X, Y)$ has copula $\textrm{C}$, if $\textrm{C}$ is a bivariate distribution function with uniform marginals and 
\begin{equation}
    \P(X \le x, \, Y\le y) = \textrm{C} \left( F_X(x), \, F_Y(y)\right)\,,
\end{equation}
where $F_X$ and $F_Y$ are the (marginal) distribution functions of $X$ and $Y$, respectively.
\end{definition}  

We denote the set containing all square integrable and $\F_T$-measurable random variables that have copula $\Co$ with the benchmark's terminal wealth $X_T^\delta$ by
\begin{equation}
    \C^\delta:=\{\; X \in \Lp(\P,\Omega)\; |\;  (X,X_T^\delta) \,\text{ has copula } \Co \text{ and } X \in \F_T\;\}.
\end{equation}

\begin{definition}
The $2$-Wasserstein distance (which we simply call the Wasserstein distance) between two random variables $Y_1$ and $Y_2$ is defined as 
\begin{equation}
d_2[Y_1\,,\, Y_2] 
    := \inf_{\chi\in\Pi(F_{Y_1},F_{Y_2})} \left\{\,\left(\int_{\R^2}|y_1-y_2|^2\,\chi(dy_1,dy_2)\right)^{\frac12}\, \right\},
\end{equation}%
where $\Pi(F_{Y_1},F_{Y_2})$ denotes the set of all bivariate probability measures  with marginal distributions $F_{Y_1}$ and $F_{Y_2}$. 
\end{definition}
The Wasserstein distance minimises the cost of transporting the distribution $F_{Y_1}$ to $F_{Y_2}$ using all possible couplings (bivariate distributions) with  fixed marginals and quadratic costs\new{, we refer to \cite{Santambrogio2015book} for a discussion on the Wasserstein distance}. One of the key reasons the Wasserstein distance is widely used to quantify model uncertainty is that it allows comparisons with distributions with differing support, thus includes comparisons to empirical distributions. Further, it is symmetric and forms a metric on the space of probability measures. \cite{dall1956SNS} shows that, for distributions on the real line, one can write the Wasserstein distance more succinctly as
\begin{equation}
  d_2[Y_1\,,\, Y_2]   	= \left(\int_0^1 \left|\Finv_{Y_1}(u) - \Finv_{Y_2}(u) \right|^2 du\right)^{\frac12}\,\label{eqn: Wasserstein},
\end{equation}%
where $\Finv_Y$ denotes the quantile function of a random variable $Y$. As the Wasserstein distance depends only on the quantile (or distribution) functions we sometimes write $d_2(\Finv_{Y_1},\, \Finv_{Y_2})$ instead of $d_2[Y_1\,,\, Y_2] $.

While the Wasserstein distance between the optimal terminal wealth and that of the benchmark's only depends on their respective distributions, the \new{specified} copula is a state-by-state \new{preference}. Thus, we consider investors who are concerned (i) about distributional closedness to the benchmark (via the Wasserstein distance), and (ii) a state-by-state outcome (via a copula). In Section \ref{sec: all-coupla}, we discuss the optimal strategy, for investors who do not wish to specify a copula but still aim to be close to the benchmark in distribution.

\subsection{Beating the Benchmark}

Equipped with a benchmark strategy $\delta$, the investor aims to construct an optimal strategy $\pi \in \A$ whose terminal wealth $X^\pi_T$ is such that the pair $(X_T^\pi\,,\, X_T^\delta)$ has copula $\Co$ and the Wasserstein distance between $X_T^\pi$ and $ X_T^\delta$ is bounded by a tolerance distance $\ep$, $\ep \ge 0$; i.e., $d_2[X_T^\pi, X_T^\delta] \le \ep$. Moreover, the investor requires that the optimal strategy's initial cost be at most that of the benchmark's and that its terminal wealth maximally improves upon an a priori chosen risk measure. Here, we consider the  class of distortion risk measures \citep{Yaari1987Economitrica}.
\begin{definition}
A distortion risk measure applied to a random variable $Y$ is 
\begin{equation}
\label{eqn: RM}
    \RM[Y] := - \E[\; Y \, \gamma\left(F_Y(Y)\right)\;]
     = - \int_0^1 \Finv_Y(u) \,\gamma(u) \, du\,, 
\end{equation}
where $\gamma \colon (0,1) \to \R_+$ is a distortion weight function such that $\int_0^1 \gamma(u) \, du = 1$. 
\end{definition}
For technical reasons we restrict our consideration to $\gamma\in\Lp([0,1])$. Analogous to our treatment of the Wasserstein distance, we write $\RM[ \cdot]$ when the risk measure is applied to random variables and $\RM( \cdot)$ when it is evaluated on quantile functions. Distortion risk measures were first introduced by \cite{Yaari1987Economitrica} and include a wide range of risk measures used in financial risk management and behavioural economics. Distortion risk measures are law-invariant, comonotone-additive, and if the distortion weight function $\gamma$ is decreasing they are coherent risk measures; see \cite{Artzner1999MF} for a discussion on properties of coherent risk measures. Moreover, the class of law-invariant, comonotone-additive, and coherent risk measures is spanned by all decreasing distortion weight functions $\gamma$ \citep{Kusuoka2001AME}. The class of distortion risk measures includes one of the most widely used risk measures in financial risk management, the TVaR at level $\alpha \in(0,1]$, with $\gamma(u) = \frac{1}{\alpha}\Id_{\{u \le \alpha\}}$ (also called Expected Shortfall); see \cite{Acerbi2002JBF}. Distortion risk measures can be used to represent investors who overweight small and underweight large probabilities through what are known as IS risk measures. These probability distortions, as introduced in \cite{Tversky1992JRU}, correspond to the case in which $\gamma$ is decreasing on $(0, u^*]$ and increasing on $( u^*, 1)$, for some $u^* \in (0,1)$. The description ``inverse S-Shaped'' pertains to the shape of the integral of $\gamma$, which does indeed have an inverse S-shape. 

Henceforth, we use risk measures to denote distortion risk measures. For the running example in this paper, we use the $\alpha$-$\beta$ risk measure -- a simplified IS risk measure detailed in the example below. We investigate the optimal strategy for an investor with IS risk measure, in Section \ref{sec: ex-comparison-optimal-strategies}.

\noindent
\begin{example}[$\alpha$-$\beta$ risk measure]
\label{ex: alpha-beta-rm}
\begin{small}
We define the $\alpha$-$\beta$ risk measure by
\begin{equation}\label{eqn:gamma}
    \gamma(u) = \tfrac{1}{\eta}\,\left( p\,\Id_{\{u\le \alpha\}} + (1-p)\,\Id_{\{u>\beta\}}\right),
\end{equation}
with normalising constant $\eta=p\,\alpha+(1-p)\,(1-\beta)$, $0<\alpha\le\beta<1$, and $p\in[0,1]$. This parametric family contains several notable risk measures as special cases:
\begin{itemize}

    \item[-] For $p = 1$, \eqref{eqn:gamma}  reduces to the TVaR at level $\alpha$. 

    \item[-] For $p\in(\frac{1}{2},1]$, \eqref{eqn:gamma} emphasises losses relative to gains.

    \item[-] For $p\in[0,\frac{1}{2})$, \eqref{eqn:gamma}  emphasises gains relative to losses.

    \item[-] For $p=0$, \eqref{eqn:gamma}  reduces to the negative of the conditional upper tail expectation (UTE) at level $\beta$. 

    \item[-] For $p\in(\frac{1}{2},1]$ and $\alpha=\beta$, \eqref{eqn:gamma} reduces to $\gamma(u) = \kappa\,\left(\frac{1}{\alpha}\Id_{\{u\le\alpha\}} +
\lambda\right)$, where $\kappa=\frac{(2p-1)\alpha}{\eta}$ and $\lambda=\frac{1-p}{(2p-1)\alpha}$, and the risk measure is equivalent to $\RM[Y]=\kappa\left(\TVaR_\alpha[Y] -\lambda\,\E[Y]\right)$. Thus, this risk measure corresponds to an investor who aims to minimise TVaR while maximising expected wealth.
\end{itemize} 
\vspace{1em}
\end{small}
\end{example}

Next, we provide a formal definition of the investor's optimisation problem and the corresponding class of optimal strategies they seek.
\begin{definition}[Optimal Strategy]
\label{def:optimal-strategy}
An \emph{optimal strategy} is a strategy $\pi\in\A$ that satisfies
\begin{enumerate}[label=(\roman*)]
    \item the budget constraint $\E[\, \varsigma_T\, X_T^\pi\,] \le X_0^\delta=\E[\, \varsigma_T\,X_T^\delta\,]$,
    \item the pair $(X_T^\pi,X_T^\delta)$ has copula $\Co$,  
    \item its terminal wealth lies within an $\ep$-Wasserstein ball around $X_T^\delta$, that is $d_2 [X_T^\pi\,, \, X_T^\delta] \le \ep$, and  
    \item $X_T^\pi$ maximally improves upon a risk measure $\RM$. 
\end{enumerate}   
Thus, the investor aims to construct a strategy $\pist$ that attains:
\begin{equation}\label{eqn: main-formulation}
\inf_{\pi\in\A}
 \RM[X_T^\pi]\, \qquad \text{subject to }\quad d_2 [X_T^\pi \,, \, X_T^\delta] \le \ep,\quad X_T^\pi\in \mathcal{C}^\delta,  \quad \text{and} \quad
\E[\, \varsigma_T\,X_T^\pi\,]\le X_0^\delta. \tag{P}
\end{equation}
\end{definition}
\new{
Of particular note is that there always exists a strategy that satisfies properties (i), (ii), and (iii) of Definition \ref{def:optimal-strategy} but is distinct from the benchmark strategy. For a complete construction of such strategies, see Appendix \ref{sec:distinct_strategies}.
}


The following theorem is based on a collection of the results in later sections, and illustrates that our problem formulation is indeed well-posed.
\begin{theorem}[Existence and Uniqueness]
\label{thm:P-existence} 
The following properties hold:
\begin{enumerate}[label=(\roman*)]

\item There exists a solution to \eqref{eqn: main-formulation}. 

\item There exists an admissible strategy $\pist$ that attains the infimum in \eqref{eqn: main-formulation}.

\item $\pist$ satisfies $d_2[X^{\pist}_T\,,\,X_T^\delta]=\ep$ and $\E[\, \varsigma_T\,X^{\pist}_T\,] \le X_0^\delta$.

\item $\pist$ satisfies $\RM[X_T^\pist] \le \RM[X_T^\delta]$.

\item The solution is unique, if the solution to \eqref{eqn: main-formulation} without the budget and copula constraints has strictly smaller risk measure than the solution to the original optimisation problem \eqref{eqn: main-formulation}.
\end{enumerate}
\end{theorem}

 \begin{figure}
    \centering
    \includegraphics[width=0.4\textwidth]{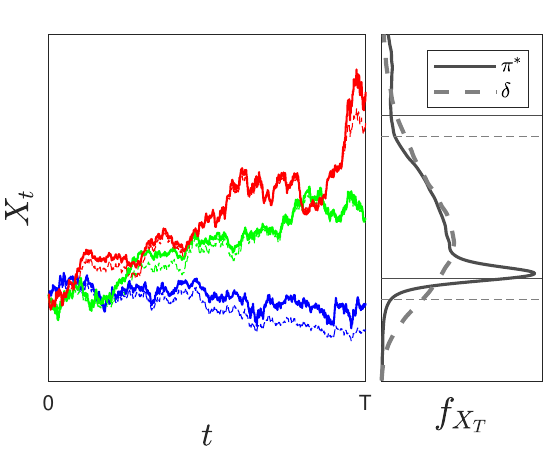}
    \caption{Optimal strategy $\pi^*$ constructed from a benchmark strategy $\delta$. Left: Three sample paths (solid line $\pi^*$, dashed line $\delta$). Right: Densities of the terminal wealth of the optimal strategy (solid line) and the benchmark (dashed line). Horizontal lines indicate the 10\% and 90\% quantiles for each distribution.}
    \label{fig:optimal-illustration}
\end{figure}
Figure \ref{fig:optimal-illustration}, shows a sketch of a benchmark strategy together with the corresponding optimal strategy $\pis$ under three scenarios (left panel), and the corresponding terminal wealth densities (right panel). In the right panel, the dashed lines represent the benchmark strategy, while the solid lines represent the optimal strategy. We also show the lower- and upper-tail quantiles of the terminal wealth distributions in the right panel. As the optimal strategy is a solution to \eqref{eqn: main-formulation}, it satisfies all constraints and, in this example, the optimal portfolio has better lower-tail risk and better upper-tail performance.

To derive the optimal strategy that solves optimisation problem \eqref{eqn: main-formulation}, we proceed via the following steps. We observe that the constraints in the optimisation problem \eqref{eqn: main-formulation} only pertain to alternative strategy's terminal wealth random variable. Thus, we split \eqref{eqn: main-formulation} into (i) optimisation problem \eqref{eqn: main-formulation-Pprime} -- an optimisation over random variables, and (ii) finding the optimal strategy that replicates the optimal terminal wealth random variable. As (ii) can be obtained via standard replication arguments, we omit this part. To derive the optimal terminal wealth random variable, we transform the optimisation over random variables to one over quantile functions \eqref{eqn: main-formulation-Pdagger}. Since the optimisation over quantile functions is not strictly convex we utilise a regularisation, problem \eqref{eqn: main-formulation-Pdagger-c}, to prove that indeed the optimal quantile function is unique. Once the optimal quantile function is derived, we obtain the optimal terminal wealth random variable via Corollary \ref{cor: solution-to-Pdagger-c}. In Section \ref{sec:simualtion-approach}, we provide an algorithm for calculating the optimal terminal wealth random variable using simulations alone.

\section{The Optimal Quantile Function}
\label{subsec: opt quant}

In problem \eqref{eqn: main-formulation}, we observe that the objective functional and all constraints pertain only to the terminal $\F_T$-measurable random variable $X_T^\pi$. Hence, solving problem \eqref{eqn: main-formulation} is equivalent to first deriving the optimal terminal random variable, that is
\begin{equation}
\label{eqn: main-formulation-Pprime}
\tag{$P^\prime$}
\inf_{X\in\Lp(\P,\Omega)}
 \RM[X]\,\quad 
 \text{subject to} \quad
 d_2 [X \,, \, X_T^\delta] \le \ep,\quad X\in \C^\delta,  \quad \text{and} \quad
\E[\, \varsigma_T\, X\,]\le X_0^\delta,
\end{equation}
and then constructing a strategy $\pi\in\A$ that attains $X_T^\pi=X$ $\P$-a.s. As $X\in\C^\delta$, it is $\F_T$-measurable. Market completeness, moreover, ensures that a unique strategy that replicates $X$ exists.

We next  derive the unique optimal terminal wealth, i.e., solve problem \eqref{eqn: main-formulation-Pprime}, the construction of the unique optimal strategy follows using standard replication arguments and is omitted for brevity.
 

To solve problem \eqref{eqn: main-formulation-Pprime}, we rewrite it as an optimisation problem over quantile functions. This reformulation is connected to the copula constraint and leads to a characterisation of the optimal terminal quantile function via an isotonic projection and, moreover, of the optimal terminal random variable. In Section \ref{sec: all-coupla}, we relax the assumption that the benchmark and the optimal terminal wealth are coupled via a specific copula, in that we consider the optimisation problem without a copula constraint and provide its corresponding solution.  
 
The quantile reformulation provides an optimisation problem with a convex objective function (the risk measure) and a convex constraint set. The objective function of problem \eqref{eqn: main-formulation-Pprime} is not convex, as the risk measures considered in this paper are not necessarily convex on the space of random variables. Whenever $\gamma$ is not decreasing, for example IS and the (generic) $\alpha$-$\beta$ risk measures, then the risk measure is not convex on the space of random variables.

To reformulate problem \eqref{eqn: main-formulation-Pprime}, we write its three constraints -- the Wasserstein, the copula, and the budget constraint -- in terms of quantile functions. First, we observe that the Wasserstein distance $d_2[X,\,X_T^\delta] = d_2(\Finv_X,\,\Finv)$  depends only on the corresponding quantile functions. Thus, we define the set of quantile functions that lie within an $\ep$-Wasserstein ball around the quantile function of the benchmark's terminal wealth $\Finv$ as
\begin{equation}
\oM_\ep := \left\{ g \in \oM ~\left|~ \int_0^1 \left(g(u)-\Finv(u)\right)^2\, du\, \le \ep^2\right.\right\},
\end{equation}%
where
\begin{equation}
\oM := \left\{\,\left. g \in \Lp([0,1])\; ~\right|~\; g \; \text{non-decreasing,  left-continuous \& } \new{g(0) = - \infty}\; \right\}
\end{equation}
is the set of square integrable quantile functions.\footnote{\new{The set $\oM$ is indeed the set of quantile functions of square-integrable random variables, where the quantile function of $X$ is defined by $
\Finv_X(u) := \left\{ y \in \R ~|~ F_X(y ) \ge y \right\}$
with the convention that $\inf\emptyset = + \infty$. This definition implies that $\Finv_X(0) = - \infty$ even for random variable with finite support for which $\lim_{u\downarrow0}\Finv_X(u)>-\infty$.}}

The next result, see e.g., Chapter 1 of \cite{Ruschendorf2013Springer}, provides a key ingredient for transforming \eqref{eqn: main-formulation-Pprime} to an optimisation problem over quantile functions.
\begin{proposition}[Quantile Representation.]
\label{prop:quantile-rep}
Let $X_T^\delta$ be a continuous random variable and set 
$U^\delta: =F(X_T^\delta)\stackrel{\P}{\sim} \U(0,1)$. Further define
\begin{equation} 
    X := \Finv_X\left(\Coninv\left(U\, |\,  U^\delta\right)\right)\,,
    \label{eqn:X-copula-Rep}
\end{equation}
where $U$ is any standard uniform random variable independent of $U^\delta$ and $\Coninv$ denotes the inverse of the conditional copula of $X^\pi$ given $X^\delta$. 
Then $X$ has distribution $F_X$ under $\P$ and $(X,X_T^\delta)$ has copula $\Co$. Moreover, any random variable $Y$ with $\P$-distribution $F_X$  and such that $(Y,X_T^\delta)$ has copula $\Co$ has to be of the form \eqref{eqn:X-copula-Rep}.
Therefore, the following representation holds
\begin{equation}
    \C^\delta = \left\{\left.\,g\left(\Coninv\left(U\, |\,  U^\delta\right)\right)\;\right|\; g\in\oM \,,\;\; U 
    \stackrel{\P}{\sim}\U(0,1) \,,\;\; U \perp^\P \;  U^\delta\,, \text{ and } U \in \F_T\;\right\}.
    \label{eqn:CdeltaRep}
\end{equation}
\end{proposition}
Thus, the set $C^\delta$ of all random variables $X$ such that $(X, X_T^\delta)$ has copula $\Co$, can be viewed as the set of random variables generated by $U^\delta$ and a $\P$-independent uniform $U$. 

The next proposition demonstrates that we may restrict to the random variable $U$ that is counter-monotonic to the conditional SDF $\varsigma_T$ given $U^\delta$. For this we define the set of random variables
\begin{equation}
 \tilde{\C}^\delta =\left\{\,g(V)\;\left|\; g\in\oM\;\right\}\right.\,, 
 \end{equation}
where 
\begin{equation}\label{eq:V}
V := \Coninv\Big(1 - F_{\varsigma_T\,|\,U^\delta}\left(\varsigma_T\;|\; U^\delta\right)\, \Big|\,  U^\delta\Big)\,.   
\end{equation}
Indeed, $1 - F_{\varsigma_T\,|\,U^\delta}\left(\varsigma_T\;|\; U^\delta\right)\stackrel{\P}{\sim}\U(0,1)$, $\P$-independent of $U^\delta$, and $\F_T$-measurable, thus $\tilde{\C}^\delta \subset \C^\delta$.

Our construction of the random variable $V$ requires that both $X_T^\delta$ and $\varsigma_T$ have a continuous distribution function. This can be relaxed, see e.g. Definition 1.11 in \cite{Ruschendorf2013Springer}, but results in a more involved representation of $V$. For simplicity of representation, we restrict to continuous distributions. 

\begin{proposition}\label{prop:c-tilde-delta}

For any $X $ that fulfils the constraints of Problem \eqref{eqn: main-formulation-Pprime}, there exists a $g \in \oM$ such that $\tilde{X}: = g(V)$ fulfils the constraints and $\RM[\tilde{X}] \le \RM [X]$. Moreover, if $X$ lies in the interior of the Wasserstein ball, then the inequality is strict.
\end{proposition}

\noindent
\begin{example}[CoIn Copula]\label{ex:CoIn}
\begin{small}
Here, we illustrate the random variables $U^\delta$ and $V$, see Equations \eqref{eqn:X-copula-Rep} and \eqref{eq:V}. For this we define the random variable 
\begin{equation}
    \tilde{U}:=  F_{\varsigma_T\,|\,U^\delta}\left(\varsigma_T\;|\; U^\delta\right)\,,
\end{equation}
such that $V$ becomes $V = \Coninv (1-\tilde{U}\, |\,  U^\delta)$.   

We continue the SIR-CEV market model of Example \ref{ex: SIR-MM} and introduce a copula, which we call the comonotonic-independent (CoIn) copula. All parameters of the SIR-CEV market model and the choice of benchmark strategy are given in Appendix \ref{app:SIR-MM-param}. The CoIn-copula is comonotonic for $u$ larger than a threshold $u^*$ and independent below $u^*$. In particular,
\begin{subequations}
\begin{align} 
\CoIn(u, v) &= 
\tfrac{1}{u^*}\,\min(u^*,u)\min(u^*,v)+
\left(\min(u,v)-u^*\right)_+,
\\
\Con\left(v\, |\,  u\right) &= \tfrac{1}{u^*} \,\min(v,u^*)\, \mathds{1}_{\{u \le u^*\}} + \mathds{1}_{\{u>u^*\}}\,\mathds{1}_{\{u\le v\}}, \quad \text{and}
\\
\Coninv\left(x\, |\,  u\right) &=  x\,u^*\,\mathds{1}_{\{u\le u^*\}}  + 
u\,\mathds{1}_{\{u> u^*\}} \,\mathds{1}_{\{x>0\}}.
\end{align}
\end{subequations}
The CoIn-copula allows us to generate optimal payoffs that are comonotonic to the benchmark when the benchmark performs well, but are independent when it performs poorly. Figure \ref{fig:CoIn-copula} illustrates scatter plots of $U^\delta$, $\tilde{U}$, and $V$ under the CoIn-copula with $u^*=0.25$ of the SIR-CEV model. The top left panel illustrates that $U^\delta$ and $\tilde{U}$ are indeed $\P$-independent. The top middle panel is a scatter plot of the CoIn-copula which is the copula between the optimal and benchmark's terminal wealth.
\begin{figure}[h!]
    \centering
    \includegraphics[width=0.6\textwidth]{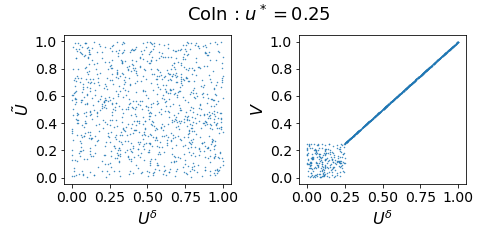}
    \caption{Illustration of the $U^\delta$, $\tilde{U}$, and $V$ under the CoIn-copula with $u^*=0.25$.}
    \label{fig:CoIn-copula}
\end{figure}
\end{small}
\end{example}

Using Proposition \ref{prop:c-tilde-delta}, we can restrict to strategies with terminal wealth in $\tilde{\C}^\delta$, which allows us to rewrite the budget constraint in terms of quantile functions. For this purpose, we define the set
\begin{equation}
\label{eqn:conditional-R-N}
    \oM^\delta := \left\{\left. g \in \oM ~\right|~  \int_0^1 g(u)\,\xi(u)\,du  \le X_0^\delta \right\},
\end{equation}
with
\begin{equation}
        \xi(u) := \E\left[\varsigma_T \; |\; V  = u\right] 
        =  \tfrac{d}{d u}\,\E[\varsigma_T \Id_{\{V \le u\}}]\,, \qquad u\in(0,1).
\end{equation}    
The function $\xi(u)$ is thus effectively the expectation of the SDF conditioned on the random variable $V = u$, which encodes the specified dependence to the benchmark strategy's terminal wealth. We make the following standing assumption:
\begin{assumption}
\label{asm:como-gamma}
If $\Co$ is the comonotonic copula, then we further assume that $\gamma(u) \neq \frac{\xi(u)}{\int_0^1 \xi(v) \, dv}$ on some nontrivial measurable subset of $[0,1]$.
\end{assumption}
This assumption prevents, under the comonotonic copula, the risk measure from being equal to the negative of the forward-neutral expectation of terminal wealth. Such a risk measure results in maximising the cost of a strategy, while simultaneously restricting the cost to be less than that of the benchmark's.

The next theorem shows that solving \eqref{eqn: main-formulation-Pprime} is equivalent to solving an alternative optimisation problem over quantile functions. 
\begin{theorem}[Quantile Formulation.]
\label{thm:quantile-formulation}
The optimisation problems
\begin{equation}
    \inf_{g\in\oM_\ep\cap \oM^\delta} \RM(g),
    \tag{$\breve{P}^\prime$}
    \label{eqn: main-formulation-Pdagger}
\end{equation}
and \eqref{eqn: main-formulation-Pprime} are equivalent in the sense that (i) given any $g^*$ that attains the infimum in \eqref{eqn: main-formulation-Pdagger}, the random variable $X^*:=g^*(V)$ attains the infimum in \eqref{eqn: main-formulation-Pprime}, and (ii) given any $X^*$ that attains the infimum in \eqref{eqn: main-formulation-Pprime}, its quantile function $\Finv_{X^*}$ attains the infimum in \eqref{eqn: main-formulation-Pdagger}.
\end{theorem}

\begin{remark}
Theorem \ref{thm:quantile-formulation} is similar in spirit to but distinct from the quantile approach discussed by \cite{he2011portfolio} where the authors consider optimal portfolios that  minimise a risk measure (subject to a budget constraint). In their work, the argument hinges on showing that the optimal terminal wealth and  the SDF are counter-monotonic. In our setting where the investor considers a benchmark, we find that the optimal terminal wealth is counter-monotonic to the conditional SDF given the benchmark. If an investor removes \new{their copula preferences, i.e. removes the copula constraint in the optimisation problem} (see Section \ref{sec: all-coupla}), and takes $\epsilon\to\infty$, i.e., there is no benchmark in the problem, then our problem reduces to the case discussed in \cite{he2011portfolio} and we recover that the optimal terminal wealth and  the SDF are indeed counter-monotonic. \cite{he2011portfolio} do not, however, consider a Wasserstein constraint, nor do they consider a benchmark strategy. As a result, under Yaari's \citeyearpar{Yaari1987Economitrica} dual model, their optimal terminal wealth distributions consist of two-point masses -- which is not satisfactory from a practical perspective. 
\end{remark}

Theorem \ref{thm:quantile-formulation} states that solutions to \eqref{eqn: main-formulation-Pprime} can be obtained by solving for optimal quantile functions $g^*$ attaining \eqref{eqn: main-formulation-Pdagger} and setting $X^* := g^*(V)$. Problem \eqref{eqn: main-formulation-Pdagger}, and hence \eqref{eqn: main-formulation-Pprime}, may not have a unique solution, as the risk measure $\RM$ is convex but not strictly convex in its argument (quantile functions). To address this issue and to show that the quantile function that attains  \eqref{eqn: main-formulation-Pprime} is indeed unique, we introduce a regularisation of \eqref{eqn: main-formulation-Pdagger}.
\begin{theorem}[Existence \& Uniqueness]
\label{thm:Existence-Uniqueness-P1p}
For all $c>0$, the solution to the regularised problem
\begin{equation}\label{eqn: main-formulation-Pdagger-c}
\inf_{g\in\oM_\ep\cap \oM^\delta}
\left( \RM(g)+ c\,\left(d_2\big(g\, , \, \Finv\big)\right)^2 \right) 
\tag{$\breve{P}^\prime_c$}
\end{equation}
exists and is unique. 
\end{theorem}
Next, we recall the concept of isotonic projections, which is fundamental to the construction of the solution to the regularised problem \eqref{eqn: main-formulation-Pdagger-c}.
\begin{definition}[Isotonic Projection]
\label{def:isotonic}
The isotonic projection $h^\uparrow$ of a function $h\in\Lp([0,1])$ is  its projection onto the set of non-decreasing (left-continuous) functions in $\Lp([0,1])$, that is,
\begin{equation}\label{eqn: isotonic projection}
h^\uparrow\, : = \argmin_{g\in \oM} \, \norm{g-h}^2.
\end{equation}
\end{definition}
We point out that the isotonic projection of $h$ is distinct from the concave envelope of $h$. Indeed, the isotonic projection is the derivative of the concave envelope of the integral of $h$. \cite{Brighi1994SIAM}[Lemma 5.1] show that the concave envelope is either the function itself or an affine function, with only countable number of such affine pieces. Therefore, the isotonic projection of a function $h$ is the increasing function $h^\uparrow$ that is constant on at most countable disjoint intervals (where the constants vary on each interval) and otherwise equals $h$. The constant parts of $h^\uparrow$  are determined such that the $\Lp([0,1])$ distance between $h$ and $h^\uparrow$ is minimised. In practice, one may approximate the isotonic projection of a function by sampling it at grid points and using one of the plethora of numerical algorithms for isotonic regression (see e.g. \cite{barlow1972statistical,tibshirani2011nearly}).

The next theorem shows that the solution to \eqref{eqn: main-formulation-Pdagger-c} is attained by a quantile function in the constraint set $\oM_\ep \cap \oM^\delta$. Moreover, the optimal quantile function is the isotonic projection of a linear combination of the benchmark terminal quantile function $\Finv$, the conditional SDF function $\xi$, and the distortion weight function $\gamma$. 
\begin{theorem}[Optimal Quantile Function]
\label{thm:isotonic-Pdagger-c}
There exists a $\bar{c}>0$ such that for all $c\in(0, \bar{c})$, the unique quantile function that attains \eqref{eqn: main-formulation-Pdagger-c} is given by
\begin{subequations}
\begin{equation}\label{eqn: pf-gstar}
    g^*_c(u):=\ell^\uparrow(u;\,\lambda_1,\lambda_2,c)\,,
\end{equation}
where 
\begin{equation}\label{eqn: ell-def}
    \ell(u;\,\lambda_1,\lambda_2,c):= \Finv(u) + \tfrac{1}{2(\lambda_1 + c)}\left(\gamma(u) -\lambda_2\,\xi(u) \right),
\end{equation}
and the Lagrange multipliers $\lambda_1, \lambda_2$, either fulfill that 
\begin{enumerate}[label = $(\roman*)$]
    \item $\lambda_1, \lambda_2>0$ are such that
\begin{equation}\label{eqn: pf-gstar-costr}
    \ep^2 = \int_0^1 \left( g^*_c(u) - \Finv(u)\right)^2\,du
    \qquad \text{and} \qquad
     X_0^\delta = \int_0^1 g^*_c(u)\,\xi(u)\,du ,
\end{equation}
or
    \item $\lambda_2 = 0$ and $\lambda_1>0 $ is such that 
    \begin{equation}
    \ep^2 = \int_0^1 \left( g^*_c(u) - \Finv(u)\right)^2\,du.
\end{equation}
\end{enumerate}
\end{subequations}
\end{theorem}
The result shows that the constraint on the Wasserstein distance is always binding, while the budget constraint may not. For the special case of the comonotonic copula, one can prove that the budget constraint is binding.

For a sufficiently small regularisation parameter $c$ (specifically for $c\in(0,\bar{c})$) Theorem \ref{thm:isotonic-Pdagger-c} shows that the Wasserstein constraint of the regularised problem \eqref{eqn: main-formulation-Pdagger-c} is binding; thus, problem \eqref{eqn: main-formulation-Pdagger-c} reduces to 
\begin{equation}
\inf_{g\in\oM_\ep\cap \oM^\delta} \RM(g)+ c\,\ep.
\end{equation}
As a consequence, we obtain the following corollary.
\begin{corollary}[Optimal Terminal Wealth]
\label{cor: solution-to-Pdagger-c}
For all $c \in (0, \bar{c})$ (where $\bar{c}$ is obtained in Theorem \ref{thm:isotonic-Pdagger-c}), the unique solution to \eqref{eqn: main-formulation-Pdagger-c}, denoted by $g^*$, is independent of $c$ and is the unique solution to \eqref{eqn: main-formulation-Pdagger}. Furthermore, 
\begin{equation}
    X^*:= g^*(V),
\end{equation}
is the unique solution to \eqref{eqn: main-formulation-Pprime}.
\end{corollary}

From Corollary \ref{cor: solution-to-Pdagger-c} and the representation in Theorem \ref{thm:isotonic-Pdagger-c} (specifically, \eqref{eqn: ell-def}), the regularisation parameter $c$ (as long as it is sufficiently small) can be absorbed by writing $\lambda_1+c\to \lambda_1$, so that we redefine
\begin{equation}\label{eq:ell}
\ell(u;\lambda_1,\lambda_2) := \Finv(u) + \tfrac{1}{2\,\lambda_1}(\gamma(u)-\lambda_2\,\xi(u))
\end{equation}
and use this version of $\ell$ throughout the remainder of the analysis.

\noindent
\begin{example}[$\alpha$-$\beta$ Risk Measure]
\begin{small}
We continue Example \ref{ex:CoIn}, and study the optimal quantile functions for $\alpha$-$\beta$ risk measures, i.e., solutions to Problem \eqref{eqn: main-formulation-Pdagger}.
\begin{figure}[h!]
    \centering
    \includegraphics[width=0.32\textwidth]{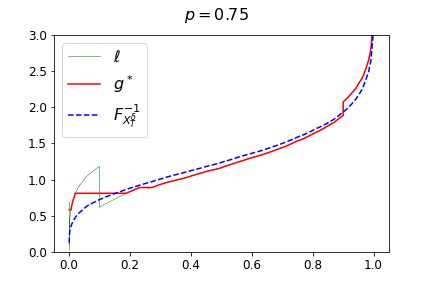}
    \includegraphics[width=0.32\textwidth]{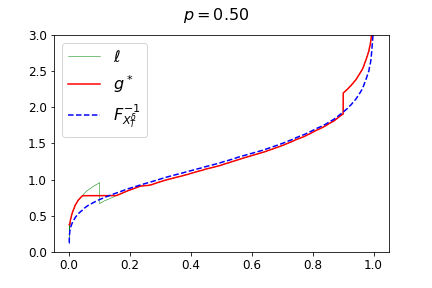}
    \includegraphics[width=0.32\textwidth]{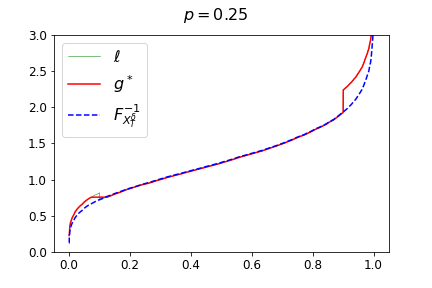}
    \caption{Optimal quantile functions compared with the benchmark's quantile function for $\ep=0.1$ and the $\alpha$-$\beta$ risk measure with $\alpha = 0.1$, $\beta = 0.9$, and various $p$. Results shown for the CoIn-copula with $u^*=0.25$.
    }
    \label{fig:inverse S-shape-G-star}
\end{figure}
Figure \ref{fig:inverse S-shape-G-star} plots the optimal quantile function $g^*$ for comparison with the benchmark's quantile function, for parameter choices $\ep=0.1$, $\alpha=0.1$, $\beta=0.9$, and three values of $p$, all under the CoIn copula with $u^* =0.25$. The green line is the function $\ell(\cdot)$ given in \eqref{eq:ell}, whose isotonic projection is equal to the optimal quantile function $g^*$. We observe in all three panels of Figure \ref{fig:inverse S-shape-G-star}, that the optimal quantile functions $g^*$ have two constant segments. The first constant segment straddles $u=\alpha$, and its length decreases as $p$ decreases. The second straddles $u= u^*$ and also decreases as $p$ decreases. While the first constant segment is induced by the risk measure, the second constant stems from the change in the dependence structure of the CoIn copula from independence to comonotonicity at $u^* = 0.25$. By contrast, a jump in $g^*$ occurs at $\beta$, and its height increases as $p$ decreases.  All of the  features describe above capture the investors' risk preferences: a large $p$ corresponds to investors protecting themselves from (downside) risk, while a small $p$ corresponds to investors emphasising gains over losses. 
\new{ Of particular note is that the  jumps at $\beta$ in the optimal quantile functions indicate that the corresponding probability densities have zero mass in the corresponding region in wealth. Thus, the benchmark and the optimal strategies lead to probability distributions that are not absolutely continuous with respect to one another, and hence a metric such as the $2$-Wasserstein distance we use here, is necessary to capture such features.}
\end{small}
\vspace{1em}
\end{example}




 
\section{Space of All Copulae}
\label{sec: all-coupla}
When the investor does not want to specify a copula, optimisation problem \eqref{eqn: main-formulation} can be amended such that $(X_T^\pis, X_T^\delta) \in \mC $, where $\mC$ is the space of all bivariate copulae. Thus, the investor considers alternative strategies that have any possible copula to the benchmark, and optimisation problem \eqref{eqn: main-formulation-Pprime} becomes
\begin{equation}
\label{eqn: main-formulation-Pall}
\tag{$P^\mC$}
\inf_{X\in\Lp(\P,\Omega)}
 \RM[X]\; 
 \text{ subject to }
 \;
 d_2 [X \,, \, X_T^\delta] \le \ep,\;\; X \text{ is } \F_T \text{-measurable}, \;\; \text{and} \;\;
\E[\, \varsigma_T\, X\,]\le X_0^\delta\,.
\end{equation}

The next two propositions illustrate how the results of this paper can be adapted to provide a solution to \eqref{eqn: main-formulation-Pall}. Recall that to solve the original optimisation problem over random variables \eqref{eqn: main-formulation-Pprime}, we first translate it to a optimisation problem over the space of quantile functions \eqref{eqn: main-formulation-Pdagger}. This quantile representation is fundamentally tied to Propositions \ref{prop:quantile-rep} and \ref{prop:c-tilde-delta}; that is the copula $\Co$. Thus, for the case when the investor does not specify a copula, we proceed as follows.
\begin{proposition}\label{prop:c-tilde-delta-all}
We define the set of random variables
\begin{equation}\label{eq:c-tilde-delta-all}
\tilde{\C}^\delta =\left\{\,g(V)\;\left|\; g\in\oM\;\right\}\right.\,,
\quad \text{where} \quad
V := 1 - F_{\varsigma_T}\left(\varsigma_T \right)\,.   
\end{equation}
Then, for any $X$ that fulfils the constraint of Problem \eqref{eqn: main-formulation-Pall}, there exists a $g \in \oM$ such that $\tilde{X}: = g(V)$ fulfils the constraints and $\RM[\tilde{X}] \le \RM [X]$. Moreover, if $X$ lies in the interior of the Wasserstein ball, then the inequality is strict.
\end{proposition}

\begin{proposition}[Solution to \eqref{eqn: main-formulation-Pall}]
\label{prop: solution-all}
The optimal terminal wealth random variable, the solution to optimisation problem \eqref{eqn: main-formulation-Pall}, is given by Corollary \ref{cor: solution-to-Pdagger-c} with $\tilde{C}^\delta$ and $V$ defined in Equation \eqref{eq:c-tilde-delta-all}. 

In particular, Theorem \ref{thm:quantile-formulation} (the quantile formulation), Theorem \ref{thm:Existence-Uniqueness-P1p} (Existence \& Uniqueness), and Theorem \ref{thm:isotonic-Pdagger-c} (Optimal Quantile Function) hold. 
\end{proposition}


We observe from Proposition \ref{prop: solution-all}, that the optimal terminal wealth is $g^*(V) = g^*(1 - F_{\varsigma_T}(\varsigma_T))$, where $g^*$ is the optimal terminal wealth quantile function given in Theorem \ref{thm:isotonic-Pdagger-c}.
Thus, we find that the optimal terminal wealth is counter-monotonic to the SDF, as in \cite{he2011portfolio} and \cite{bernard2014explicit}, however, here, the optimal quantile function is tied to the benchmark's via the Wasserstein constraint and thus partially inherits  its structure.

\section{Choice of the Wasserstein Distance}
\label{sec: choice-of-ep}
In this section, we address the question of how an investor may choose the Wasserstein distance $\ep$ and present one approach related to an investor's tolerance of deviating from the benchmark portfolio. The approach is to choose $\ep$ such that an investor's lower and upper bound on the mean and standard deviation, respectively, are not violated. 

We first investigate a related and illuminating question: What are the maximal and minimal means and standard deviations of all quantile functions within the $\ep$-Wasserstein ball around the benchmark's? For this, define the minimal and maximal means of all quantiles in $\oM_\ep$ as
\begin{equation}
\ml :=\inf_{g\in\oM_\ep} \int_0^1 g(u)\,du
\quad \text{and}
\quad
\mr: =\sup_{g\in\oM_\ep} \int_0^1 g(u)\,du,
\end{equation}
and define  the minimal and maximal standard deviations in $\oM_\ep$ with fixed mean $m \in [\ml, \,\mr]$ as
\begin{equation}
\sl :=\inf_{\substack{g\in\oM_\ep\\ \int_0^1 g(u)\, du = m}} \left(\int_0^1\left( g(u) - m\right)^2\,du\right)^\frac12
\quad \text{and}
\quad
\sr: =\sup_{\substack{g\in\oM_\ep\\ \int_0^1 g(u)\, du = m}} \left(\int_0^1\left( g(u) - m\right)^2\,du\right)^\frac12.
\end{equation}

The following two propositions and the corollary show that the sets of attainable means and standard deviations within the $\ep$-Wasserstein ball are $[\m-\ep\,,\,\m+\ep]$ and $[\max\{\,\s-\ep, 0\,\}\,,\,\s+\ep]$, respectively. 

\begin{proposition}
\label{prop: range-P-mean}
The minimal and maximal means that any quantile function in $\oM_\ep$ can attain are 
\begin{equation}
\ml = \m  - \ep \qquad \text{and} \qquad \mr = \m  + \ep\,,
\end{equation}
and the quantile functions attaining $\ml$ and $\mr$ are, respectively, 
\begin{equation}
\underline{g}(u) = \Finv(u)  - \ep  \qquad \text{and} \qquad \overline{g}(u) = \Finv(u)  + \ep. 
\end{equation}
\end{proposition}

\begin{proposition}
\label{prop: range-P-sd}
Let $m\in [\ml\,,\, \mr]$ be fixed, and define $\Delta m := \m - m$. Then, the minimal and maximal standard deviations that any quantile function with mean $m$ in $\oM_\ep$ can attain are
\begin{equation}\label{prop: range-P-sd-sd-values}
\sl = \max\left\{\s  - \sqrt{\,\ep^2 - (\Delta m)^2 \,} \,,\; 0\, \right\} \quad \text{and} \quad 
\sr = \s  + \sqrt{\,\ep^2 - (\Delta m)^2 \,} \, .
\end{equation}
If $\s \ge \sqrt{\,\ep^2 - (\Delta m)^2 \,} $, then the quantile function attaining $\sl>0$ is
\begin{align}\label{prop: range-P-sd-gl}
    \underline{g}(u) =\left(1 - \tfrac{\sqrt{\,\ep^2 - (\Delta m)^2 \,}}{\s}\right)  \left(\Finv(u)  - \m\right) +m\,;
\end{align}
otherwise, $\underline{g}(u) = m$, and $\underline{s} = 0$. The quantile function attaining $\overline{s}$ is  
\begin{align}\label{prop: range-P-sd-gu}
    \overline{g}(u) =\left(1 + \tfrac{\sqrt{\,\ep^2 - (\Delta m)^2 \,}}{\s}\right)  \left(\Finv(u)  - \m\right) +m\, . 
\end{align}
\end{proposition}

The range of the standard deviation of $\oM_\ep$ is given in the following corollary.
\begin{corollary}
\label{cor: min-max-std-in-oM}
    The range of the standard deviation of any quantile function in $\oM_\ep$ is $[\max\{\,\s  - \ep,0\,\}\,, \, \s  + \ep]$.
\end{corollary}
With the above results, an investor may choose the Wasserstein distance $\ep$ as follows. Suppose that the investor has a lower bound $m_l < \m$ for the mean and an upper bound for the standard deviation $s_u> \s$ that they are willing to tolerate. Then, by Proposition \ref{prop: range-P-mean} and Corollary \ref{cor: min-max-std-in-oM}, the choice $\ep=\min\{\m-m_l\,,\,s_u-\s\}$  ensures that neither of the investor's tolerances is violated.

\section{Simulation Approach}\label{sec:simualtion-approach}

In this section, we demonstrate how to obtain $g^*$ in a manner that requires only simulations of the market model and the SDF.
An implementation of this algorithm for the SIR-CEV model (which is easily adaptable to any market model one can simulate) may be found at \url{https://github.com/sebjai/Portfolio-Wasserstein-Ball}.

{
\footnotesize
\underline{Simulation Steps:}
\begin{center}
\begin{adjustwidth}{0.05\textwidth}{0.05\textwidth}
\begin{enumerate}
\item Simulate benchmark terminal wealth and SDF by:
\begin{enumerate}
        \item Simulate the Brownian motions $(W_t^1,\dots,W_t^d)_{t=0,1,\dots,T}$. We denote realisations of the simulation by $\{(w_t^{1(i)},\dots,w_t^{d(i)})_{t=0,1,\dots,T}\}_{i=1}^N$, so that the superscript ${}^{(i)}$ represents the $i^{th}$-simulation and we use this notation for all realisations of simulated random variables.
    
    \item Use $W$ to simulate asset prices $(S_t^1,\dots,S_t^d)_{t=0,1,\dots,T}$ and SDF $(\varsigma_t)_{t=0,1,\dots,T}$ by discretising the SDEs of the underlying market model.
    
    \item Use $S$ to simulate benchmark wealth path $(X_t^\delta)_{t=0,1,\dots,T}$ by discretising \eqref{eqn:dX_upsilon}.
    
\end{enumerate}

    \item Generate uniform random variables $U^\delta$, $\tilde{U} := F_{\varsigma_T|U^\delta}(\varsigma_T|U^\delta)$, and $V$ by:
    \begin{enumerate}
        \item Obtain kernel density estimate (KDE) of $F_{X_T^\delta}(x)$:
        \begin{equation}
            \hat{F}_{X_T^\delta}(x)  := \tfrac1N \sum_{i=1}^N \Phi\left(\tfrac{x-x_T^{\delta(i)}}{h_\delta}\right).
        \end{equation}
        Here, $h_\delta$ denotes a band width, and we use a Gaussian kernel, but any other kernel that respects the support of the terminal wealth random variable would be equally valid.
        
        \item Generate realisations $u^{\delta(i)}:=\hat{F}_{X_T^\delta}\left(x_T^{\delta(i)}\right)$ of $U^\delta$.
        
        \item Obtain KDE of $F_{\varsigma_T|X_T^\delta} (z|x)$:
        \begin{equation}
            \hat{F}_{\varsigma_T|X_T^\delta}(z|x) := \displaystyle\frac{\displaystyle\sum_{i=1}^N\Phi\left(\frac{z-\varsigma_T^{(i)}}{h_\varsigma}\right)\,\phi\left(\frac{x-x^{\delta(i)}_T}{h_x}\right)}{\displaystyle\sum_{i=1}^N\phi\left(\frac{x-x^{\delta(i)}_T}{h_x}\right)} 
        \end{equation}
        
        \item Generate realisations $\tilde{u}^{(i)} :=\hat{F}_{\varsigma_T|U^\delta}(\varsigma_T^{(i)}|u^{\delta(i)}) = \hat{F}_{\varsigma_T|X^\delta_T}(\varsigma_T^{(i)}|x^{\delta(i)}_T)$ of $\tilde{U}$. 
        

        
        \item Generate realisations $v^{(i)}:=\Coninv(1-\tilde{u}^{(i)}|u^{\delta(i)})$ of $V$.
        
    \end{enumerate}
    
    \item Obtain an estimate of $\xi(u)$ by 
    \begin{enumerate}
        \item  noting that
        \begin{align}
            \xi(u) &= \E[\varsigma_T|V=u] = \frac{d}{du}\E[\varsigma_T\Id_{V\le u}]
            =\E[\varsigma_T] \;\frac{d}{du}\E\left[\frac{\varsigma_T}{\E[\varsigma_T]}\Id_{V\le u}\right] 
            = \E[\varsigma_T] \frac{d}{du} \Q_T(V\le u)\,,
        \end{align}
        where $\Q_T$ is the probability measure induced by the Radon-Nikodym derivative $\frac{d\Q_T}{d\P}:=\frac{\varsigma_T}{\E[\varsigma_T]}$.
        
        \item Obtain KDE of $\Q_T(V\le v)$:
    \begin{equation}
        \hat{\Q}_T(V\le v) := \tfrac1N \sum_{i=1}^N \left(\frac{\varsigma_T^{(i)}}{\frac1N \sum_{j=1}^N\varsigma_T^{(j)}} \right) \Phi\left(\frac{v-v^{(i)}}{h_v}\right)\,,
    \end{equation}
    where either an logit-transformation of $V$ or a kernel with support on $[0,1]$ should be used.
    
    \item An estimate of $\xi(u)$ is therefore
    \begin{equation}
        \hat{\xi}(u):= \tfrac1N \sum_{i=1}^N \varsigma_T^{(i)} \;\tfrac{1}{h_v}\phi\left(\frac{v-v^{(i)}}{h_v}\right)\,.
    \end{equation}
    \end{enumerate}
 
    \item Let $0<u_1<u_2<\cdots<u_M<1$ denote a partition of $(0,1)$. Evaluate $\hat\xi(u_i)$ and $\gamma(u_i)$. Invert $\hat{F}_{X_T^\delta}(x_i)=u_i$ to obtain an estimate of $\Finv_{X_T^\delta}(u_i)$. 
    
    \item Use Theorem \ref{thm:isotonic-Pdagger-c}, and specifically Equation \eqref{eqn: pf-gstar}, to estimate $g^*(u)$  and obtain the Lagrange multipliers that enforce the constraints:
    \begin{enumerate}
        \item Pick $\lambda_2\ge 0$
        \begin{enumerate}
            \item Pick $\lambda_1>0$
            \item compute isotonic regression of $\ell(u;\lambda_1,\lambda_2):=\Finv_{X_T^\delta}(u) + \frac{1}{2\lambda_1}(\gamma(u)-\lambda_2\,\xi(u))$ using a numerical method (e.g., in Python the IsotonicRegression function from the sklearn.isotonic package)
            \item compute the Wasserstein distance $\left(\int_0^1 (\ell^\uparrow(u)-\Finv_{X_T^\delta}(u))^2\right)^\frac{1}{2}\,du$
            \item repeat from step i. until Wasserstein distance is satisfied
        \end{enumerate}
        \item compute cost of strategy $c[\ell^\uparrow]:=\int_0^1 \ell^\uparrow(u)\,\xi(u)\,du$
        \item repeat from step (a) until either $c[\ell^\uparrow]=X_0^\delta$, or $\lambda_2=0$ (in which case $c[\ell^\uparrow]<X_0^\delta$)
        \item Let $g^*$ denote the optimal quantile function
    \end{enumerate}
    
    \item  Use scenarios of $V$ from step 2. and $g^*$ from step 5. to generate scenarios $X_T^{\pi^*}:=g^*(V)$ 
    
    \item Use any standard method to obtain the replicating strategy for $X_T^{\pi^*}$, e.g., solving a PDE or Monte Carlo regression methods, or more modern approaches such as \cite{Ludkovski2020WP} who use Gaussian processes to estimate hedging strategies.
\end{enumerate}
\end{adjustwidth}
\end{center}
}

As a test of the simulation approach, we derive the semi-analytical form of $\xi(u)$ in a Geometric Brownian motion (GBM) market model for a constant benchmark and compare it with the proposed simulation approach.

\noindent
\begin{example}[GBM Market-Model \& Constant Benchmark]
\label{ex:GBM-CoIn}
\begin{small}
Under the GBM market model the $\Q_T$-distribution of $V$, and thus also $\xi(u)$, admit a semi-analytical form given in the next proposition.

\begin{proposition}\label{prop:GBM-CoIn}
For the GBM market model and a constant benchmark we have
\begin{align}
U^\delta&=\Phi\left(\tfrac{1}{\sqrt{T}} \vartheta^\intercal W_T\right)\,,
\\
\tilde{U}:=F_{\varsigma_T|U^\delta}(\varsigma_T|U^\delta)&=\Phi\left(\tfrac{1}{\sqrt{T}} \theta^\intercal W_T\right)\,,
\end{align}
which are $\P$-independent uniforms, $\vartheta:=\frac{\eta}{\sqrt{\eta^\intercal\rho\eta}}$, and 
$\theta:=\frac{-\lambda+\frac{\lambda^\intercal\rho\eta}{\eta^\intercal\rho\eta}\eta}
{
\sqrt{
\left(
\lambda^\intercal\rho\lambda-\frac{(\lambda^\intercal\rho\eta)^2}{\eta^\intercal\rho\eta}
\right)
}
}$. 
Moreover, for any copula $\Co(\cdot,\cdot)$ it holds that $\xi(u)=e^{-r\,T}\frac{d}{dv}\Q(V\le v)$ and
\begin{equation}
\Q_T(V\le v) =     
\int_0^1 
\Phi\left( \Phi^{-1}\left(1-\Con\left(v\, |\,  u\right) \right)+\theta^\intercal\,\rho\,\lambda\,\sqrt{T}\right) 
\;
d\Phi\!\!\left( \Phi^{-1}(u) +\frac{\eta^\intercal\rho\lambda}{\Psi}\,\sqrt{T}\right) \,,
\label{eqn:Q_V_GBM}
\end{equation}
\end{proposition}
where $\lambda := \rho^{-1} \tfrac{\mu - r}{\sigma}$, 
$\eta:=(\delta^1\sigma^1,\delta^2\sigma^2,\dots,\delta^d\sigma^d)^\intercal$, and $ \Psi^2 :=\eta^\intercal\rho\eta$.

\begin{figure}[!t] 
\begin{center}
    \centering
    \includegraphics[width=0.3\textwidth]{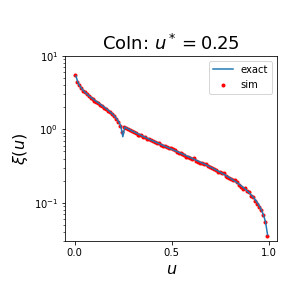}
    \quad
    \includegraphics[width=0.3\textwidth]{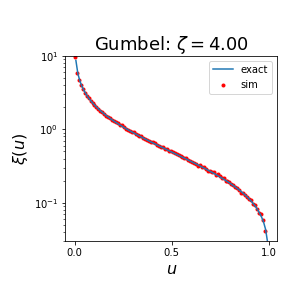}
\end{center}
    \caption{Comparison of $\xi(u)$ calculated using the semi-analytical formula (blue line) and via the simulation approach (red dots); CoIn copula (left) and Gumbel copula (right).
    \label{fig:xi-CoIn-Gumbel}}
\end{figure}

Figure \ref{fig:xi-CoIn-Gumbel} plots the function $\xi(u)$ using the semi-analytical formula, solid blue line, and the simulation approach, red dots. The left panel shows $\xi(u)$ for the CoIn copula and the right panel for the Gumbel copula. We can clearly see that the simulation approach is capable of capturing the shape of $\xi(u)$. Note that the kink in the left panel is at $u = u^* = 0.25$, induced by the change in dependence structure of the CoIn copula.
\end{small}
\end{example}

\section{Simulations \& Examples}\label{sec: simulation-example}


In this section, we explore some of the numerical consequences of our optimal portfolio construction. Throughout this section and the numerical implementation, we consider the SIR-CEV market model introduced in Example \ref{ex: SIR-MM} and benchmark strategies $\delta$ that are constant. The specific SIR-CEV market model parameters are given in Appendix \ref{app:SIR-MM-param} and $\delta=(0.2,0.6,0.1)$. Note that the numerical complexity of the simulation approach grows only linearly in the number of assets, however, we choose to run the simulations with two risky assets, plus the bond, for illustrative purpose. 

In Section \ref{sec: ex-comparison-optimal-strategies}, we discuss the optimal terminal wealth for different $\alpha$-$\beta$ and IS risk measures and highlight how investors optimal terminal wealth differs from the benchmark's. Section \ref{sec: ex-choice-of-copula} compares the optimal terminal wealth under different copula assumption and contrast it with the optimal terminal wealth when the investor does no specify a copula, that is the solution to optimisation problem \eqref{eqn: main-formulation-Pall}.


\subsection{Investor's Risk Measures}\label{sec: ex-comparison-optimal-strategies}

Here, we compare the optimal strategies of investors with different risk measures. We consider an investor who optimises $\TVaR_\alpha$ (i.e., the $\alpha$-$\beta$ risk measure with $\alpha = 0.1$ and $p = 1$), an investor who maximises $\UTE_\beta$ (i.e., the $\alpha$-$\beta$ risk measure with $\beta = 0.9$ and $p = 0$), an investor minimising a linear combination of $\TVaR_\alpha$ and (negative) expectation  ($\TVaR$ \& E) (i.e., the $\alpha$-$\beta$ risk measure with $\alpha = \beta = 0.1$ and $p = 0.75$), and an investor with an IS risk measure. We consider the IS risk measure with weight function 
\begin{equation}
    \gamma(u) = \frac{d}{du}\;\frac{u^q}{(u^q+(1-u)^q)^{\frac1q}}\,, \qquad q\in(0,1)\,,
\end{equation}
which corresponds to linear utility in the CPT framework, and use $q = 0.6$ a typical value as estimated by \cite{Tversky1992JRU}. 

Table \ref{tab:benchmark-example} reports the different risk measures introduced above of the benchmarks terminal wealth, along with the mean and standard deviation (std.dev.) of the benchmark's return. We further calculate the Gain-Loss-Ratio (GLR), using the expected wealth of the benchmark strategy as the reference point  \citep{de2015bang}; this widely used metric is explicitly defined as
\begin{equation}
    GLR = \displaystyle\frac{
    \displaystyle\E\left[\, \left(X_T^\pis/X_0^\pis - \E\left[X_T^\delta/X_0^\delta\right]\right)_+ \,\right]
    }{
    \displaystyle\E\left[ \, \left(\E\left[X_T^\delta/X_0^\delta\right]-X_T^\pis/X_0^\pis \right)_+\,\right]
    },
\end{equation}
where we use the benchmark's expected return as the cutoff value.
\begin{table}[htbp]
  \centering
  \caption{Various risk measures and statistics of the benchmark strategy.}
  \footnotesize
    \begin{tabular}{ccccccc}
    \addlinespace
    \toprule\toprule
    $\TVaR_{0.1}$  & $\UTE_{0.9}$   & $\TVaR_{0.1}$ \& E & IS    & $mean$ &   $std.dev.$ & $GLR$ \\[0.25em]
    -0.58 & -2.25 & -1.17 &  -1.53 &  28.9\% & 48.2\% & 1.000 \\
    \bottomrule\bottomrule
    \end{tabular}
  \label{tab:benchmark-example}
\end{table}

Table \ref{tab:relative-change-example} presents a comparison of the optimal and benchmark strategies for different Wasserstein distances $\ep$. We observe that the larger $\ep$, the smaller the risk measure of the optimal portfolio, as larger $\ep$'s provide more flexibility in the choice of alternate strategies. The behaviour of the mean, std.dev. (of returns), and GLR of the optimal portfolio on the Wasserstein distance $\ep$, however, depends on the particular risk measure being optimised. For TVaR, all three additional measures decrease with increasing $\ep$. This occurs as the investor aims to reduce the risk as measured by the left tail, and a larger Wasserstein distance provides more flexibility to deviate from the benchmark's distribution, but doing so reduces gains as well. The opposite happens for the UTE risk measure, i.e., increasing $\ep$ leads to increases in mean, std.dev., and GLR. For TVaR \& E, we observe that mean, std.dev., and GLR all decrease except for the largest $\ep$ where they all increase. This stems from the extra freedom that a large Wasserstein ball provides, indeed while for small $\ep$ the optimal portfolio favors reducing the risk associated with the left tail, for larger $\ep$, the combination of the TVaR and expectation may be decreased further by increasing the mean. A finer grid in $\ep$ (not reported here) shows that a similar, albeit different, pattern can be observed for the IS risk measure, where the mean and GLR decrease, with increasing $\ep$, but the std.dev. increases for large $\ep$.
\begin{table} [h] 
  \centering
  \caption{Comparisons of the optimal and  benchmark strategies for different $\ep$ values. The first column indicates the investors' risk measure, where TVaR \& E corresponds to an investor minimising TVaR while maximising the expected wealth and IS corresponds to the IS risk measure.}
    \footnotesize
    \begin{tabular}{rrrrrrrrrr}
    \addlinespace
    \toprule
          & $\ep^2$ &       & $\RM(G^*)$ &       & $mean$ &       & $std.dev.$ &       & $GLR$ \\
    \midrule
    \multirow{4}{*}{TVaR} & $10^{-5}$ &       & -0.59 &       & 33.1\% &       & 49.5\% &       & 1.24 \\
     & $10^{-4}$ &       & -0.61 &       & 32.6\% &       & 49.0\% &       & 1.21 \\
     & $10^{-3}$ &       & -0.67 &       & 31.1\% &       & 47.1\% &       & 1.13 \\
     & $10^{-2}$ &       & -0.83 &       & 27.7\% &       & 43.7\% &       & 0.93 \\
     \midrule
    \multirow{4}{*}{UTE} & $10^{-5}$ &       & -2.26 &       & 33.3\% &       & 50.0\% &       & 1.25 \\
     & $10^{-4}$ &       & -2.28 &       & 33.4\% &       & 50.4\% &       & 1.26 \\
     & $10^{-3}$ &       & -2.35 &       & 33.7\% &       & 51.7\% &       & 1.27 \\
     & $10^{-2}$ &       & -2.57 &       & 34.6\% &       & 56.0\% &       & 1.32 \\
     \midrule
    \multirow{4}{*}{TVaR \& E} & $10^{-5}$ &       & -1.17 &       & 33.1\% &       & 49.6\% &       & 1.24 \\
     & $10^{-4}$ &       & -1.18 &       & 32.8\% &       & 49.1\% &       & 1.22 \\
     & $10^{-3}$ &       & -1.21 &       & 31.9\% &       & 47.5\% &       & 1.17 \\
     & $10^{-2}$ &       & -1.23 &       & 34.9\% &       & 51.7\% &       & 1.35 \\
     \midrule
    \multirow{4}{*}{IS} & $10^{-5}$ &       & -1.54 &       & 33.2\% &       & 49.8\% &       & 1.24 \\
     & $10^{-4}$ &       & -1.55 &       & 32.8\% &       & 49.6\% &       & 1.22 \\
     & $10^{-3}$ &       & -1.58 &       & 31.4\% &       & 48.7\% &       & 1.14 \\
     & $10^{-2}$ &       & -1.69 &       & 29.8\% &       & 51.1\% &       & 1.05 \\
    \bottomrule
    \end{tabular}    
  \label{tab:relative-change-example}
\end{table}

\begin{figure}
    \centering
    \includegraphics[width=0.7\textwidth]{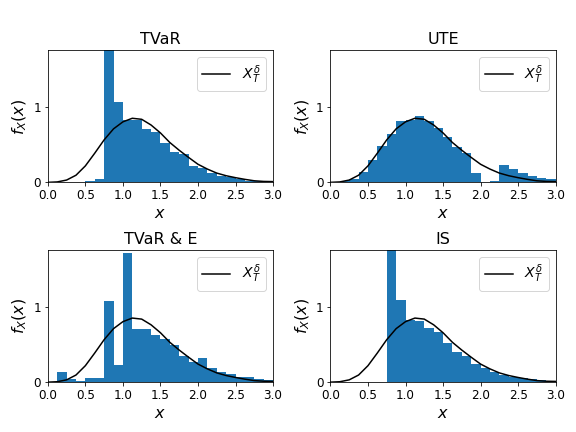}
    \caption{Comparisons of the optimal and benchmark's terminal wealth for an investor with TVaR, UTE, TVaR and expectation and,  IS risk measure. The Wasserstein distance is $\ep = 0.1$ and we consider the CoIn copula with $u^* = 0.25$.}
    \label{fig:Opt-Terminal-density-compare}
\end{figure}

Next, we discuss the specific case of $\ep = 0.1$ illustrated in Figure \ref{fig:Opt-Terminal-density-compare}. We observe that the $\TVaR$ investor's optimal terminal wealth (top left panel) has less mass in the left tail than the benchmark's, indicating a smaller $\TVaR$. Indeed, the optimal strategy results in a relative improvement of 43\% of the $\TVaR$; see Table \ref{tab:relative-change-example}. Moreover, the probability mass of the benchmark's left tail and some of its right tail is shifted to a spike at $g^*(\alpha)\approx 0.84$. The investor optimising UTE (top right panel) has a substantially heavier right tail than the benchmark or any of the other strategies. Indeed, for this investor, a large portion of the probability mass is transferred from the range $(g^*(\beta^-),g^*(\beta^+))\approx(1.92,2.23)$ to the right tail.  The bottom left panel corresponds to an investor who minimises a linear combination of $\TVaR$ and expected terminal wealth. Compared to minimising the TVaR only, the optimal terminal wealth achieved by this investor has a heavier left,  an extra peak in the centre of the distribution, and a heavier right tail. Some of the probability mass in the left tail is transferred to the spike at $g^*(\alpha)\approx 0.80$ induced by the risk measure and to $g^*(u^*)\approx1.04$, which is induced by the CoIn copula where it changes from being independent to comonotonic. Finally, the optimal density for the investor with IS risk measure (bottom right panel) has little mass in the left tail. \new{Similarly to the earlier example with the $\alpha$-$\beta$ risk-measure seen in Figure \ref{fig:inverse S-shape-G-star}, the cases of TVaR, UTE, and IS all have optimal strategy terminal wealth distributions' that have a different support compared with the benchmark's. }

As discussed in Section \ref{sec: choice-of-ep}, an investor may choose the Wasserstein distance $\ep$ in accordance with their tolerance for deviating from the mean and standard deviation of the benchmark's terminal wealth. For example, an investor who wishes to maximally deviate from the mean and standard deviation of the benchmark's terminal wealth by (no more than an absolute change of) 1\% chooses $\ep = 0.01$. 


\subsection{Investor's Choice of Copula} \label{sec: ex-choice-of-copula}
In this section, we fix the $\alpha$-$\beta$ risk measure with parameters $\alpha = 0.1$, $\beta = 0.9$, and $p = 0.75$, and compare the optimal terminal wealth random variable for the CoIn copula, the Gumbel copula, and no specified copula, that is the solution to optimisation problem \eqref{eqn: main-formulation-Pall}. 
\begin{figure}
    \centering
    \includegraphics[width=0.6\textwidth]{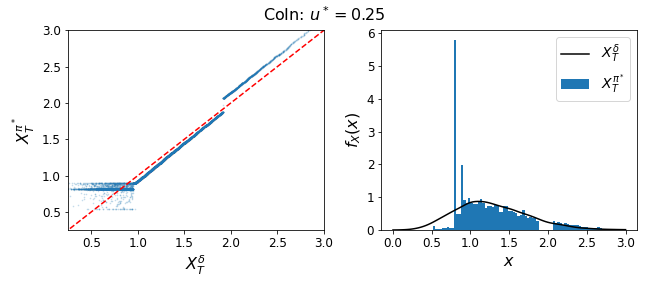}
    \\
    \includegraphics[width=0.6\textwidth]{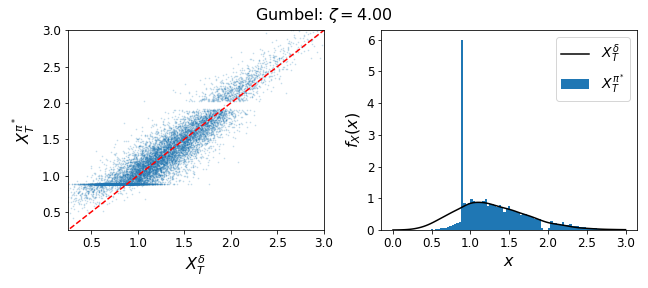}
    \\
    \includegraphics[width=0.6\textwidth]{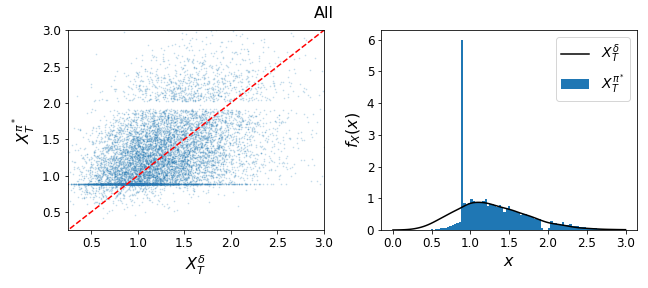}
    \caption{Scatter plot of terminal wealth of benchmark and optimal portfolio and histogram of terminal wealth using three different copulae and $\ep = 0.1$.}
    \label{fig:g(v)-vs-X^delta}
\end{figure}

Figure \ref{fig:g(v)-vs-X^delta} displays for each of the three copulae, on the lhs, a scatterplot of the optimal terminal wealth random variable $X_T^\pis = g^*(V)$ against that of the benchmark's $X_T^\delta$, and, on the rhs, the corresponding histograms. The top panel illustrates the CoIn copula with $u^* = 0.25$ and we observe that $X_T^\pis$ and $X_T^\delta$ are indeed comonotonic for realisations of $X_T^\pis $ larger than $\Finv(u^*)$ and independent below. The second panel displays the Gumbel copula, which possesses upper tail dependence but no lower tail dependence. For the definition and properties of tail dependence, see e.g., \cite[Chapter 7]{Mcneil2015book}. The upper tail dependence can be observed in the concentration of the simulated points of $(X_T^\pis, X_T^\delta)$ in the upper right corner, indicating that the investor wishes to be more strongly dependent when the benchmark does well compared to when the benchmark does poorly (zero lower tail dependence). The bottom panel displays the optimal terminal wealth when the investor does not specify a copula (optimisation problem \eqref{eqn: main-formulation-Pall}). This scatter plot clearly shows that without a prespecified copula, there is little dependence between the optimal terminal wealth random variable $X_T^\pis$ and that of the benchmark's $X_T^\delta$, compared with the Gumbel and CoIn copulae. However, $X_T^\pis$ and $X_T^\delta$ are not independent. 

The pdfs in the figure all contain (at least) one spike and a single trough induced by the $\alpha$-$\beta$ risk measure. The left-most spike corresponds to the flat part of the optimal quantile function that straddles $u=\alpha$. The trough corresponds to the jump in the quantile function at $u=\beta$. The investor's risk measure thus induces a contraction of the optimal strategy's terminal wealth distribution between $g^*(\beta^-)$  and $g^*(\beta^+)$, causing the corresponding probability mass to be transferred to the right tail. 

\section{Conclusions}

We solve the continuous-time dynamic portfolio optimisation problem in which an investor is seeking to improve a benchmark strategy's distortion risk measure subject to a budget constraint, such that the terminal wealth of the optimal and benchmark strategies have a prespecified copula, and the optimal strategy's terminal wealth is within an $\ep$-Wasserstein ball of the benchmark's. 

We derive the optimal terminal distribution in multiple steps: First, we split the problem into (i) an optimisation over random variables, and (ii) finding the optimal strategy. Second, we transform the problem over random variables into a constrained convex optimisation problem over quantile functions, which we solve using the concept of isotonic projection. Third, we derive the optimal terminal wealth random variable. Fourth, we  propose a simulation approach to calculate the optimal terminal wealth using only market model simulations. We provide illustrative examples of the optimal terminal wealth and  compare the results corresponding to different risk measures, including the TVaR and an IS risk measure. Furthermore, we compare the optimal terminal wealth for the CoIn copula, the Gumbel copula, and when the investor does not specify a copula.   

Our results complement those in the literature in that the Wasserstein distance (a) represents an investor's willingness to deviate from a benchmark strategy, and (b) provides optimal strategies that partially inherit the benchmark's structure, which is in stark contrast to the findings of \cite{He2011MS, he2011portfolio}, \cite{Jin2008MF}, and \cite{ruschendorf2019construction}. Moreover, \new{specifying a} copula provides investors with the ability to tie their strategy to the benchmark in a state-by-state manner.

\ACKNOWLEDGMENT{The authors thank an anonymous referee for suggesting the CoIn copula example. SJ and SP would like to acknowledge support from the Natural Sciences and Engineering Research Council of Canada (grants RGPIN-2018-05705, RGPAS-2018-522715, and DGECR-2020-00333, RGPIN-2020-04289).}

%
%
%

\begin{APPENDICES}

\footnotesize

\section{Proofs}

\subsubsection*{Proof of Theorem \ref{thm:P-existence}}
The proof is a collection of statements proven in the sequel. 

$(i)$ and $(ii)$: From Corollary \ref{cor: solution-to-Pdagger-c} there exists a random variable $X^*$ that minimises the risk measure, respects the budget constraint, $(X^*,X^\delta)$ has copula $\Co$, and $X^*$ is $\F_T$ measurable -- i.e., $X^*$ is a solution of \eqref{eqn: main-formulation-Pprime}. Since we consider complete market models, there exists a strategy $\pi \in \A$ attaining $X^*$.

$(iii)$: By Theorem \ref{thm:isotonic-Pdagger-c} the Wasserstein constraint is binding for the optimal terminal wealth and its cost is at most that of the benchmark's.

$(iv)$: See Lemma \ref{lemma:risk-measure}.

$(v)$: The proof outline is as follows: (A) we argue that  the optimal solution must have terminal wealth $g(V)$ for some quantile function $g$, where $V$ is given in \eqref{eq:V}, and (B) as the solution to problem \eqref{eqn: main-formulation-Pprime} for random variables of the form $g(V)$ is proven unique in Corollary \ref{cor: solution-to-Pdagger-c}, the result follows. All that remains is to prove part (A).

Denote by $\tilde{\pi}$ a solution to optimisation problem \eqref{eqn: main-formulation} and by $\tilde{X}$ its terminal wealth random variable with quantile function $\tilde{g}$. Let $\pi^\dagger$ be the solution to the optimisation problem
\begin{equation}\label{eqn: main-formulation-no-buget}
\inf_{\pi\in\A}
 \RM[X_T^\pi]\, \qquad \text{subject to }\quad d_2 [X_T^\pi \,, \, X_T^\delta] \le \ep\,,
 \quad \text{and }\quad
 X_T^\pi \in\F_T 
\tag{$P^{\dagger}$}
\end{equation}
and denote its terminal wealth random variable by $X^\dagger$ with corresponding quantile function  $g^\dagger$. Further, assume that $\RM[X^{\dagger}] < \RM[\tilde{X}]$ and furthermore that $\tilde{X} \neq \tilde{g}(V)$ $\P$-a.s.. Next, we construct a random variable that fulfils all constraint but has strictly smaller risk measure than $\tilde{X}$; leading to a contradiction.

For this let $w \in [0,1]$ and define the family of random variables 
\begin{equation}
    X_w := w \;\tilde{g}(V) + (1 - w) \;g^\dagger(V). 
\end{equation}
Next we show that for all $w \in [0,1]$ the random variable $X_w \in\C^\delta $ and satisfies
$d_2 [X_w \,, \, X_T^\delta] \le \ep$. Note that we can write $X_w = h(V)$ with quantile function $h(v) = w \tilde{g}(v) + (1 - w) g^\dagger(v) $. Thus, $X_w \in \tilde{\C}^\delta \subset \C^\delta$. The Wasserstein distance between $X_w$ and $X_T^\delta$ is
\begin{equation}
    d_2 [X_w \,, \, X_T^\delta] 
    \le 
    w \; d_2 [\tilde{g}(V) \,, \, X_T^\delta] + (1 - w) \;d_2 [g^\dagger(V) \,, \, X_T^\delta] 
    \le
     \ep\,,
\end{equation}
since both $\tilde{g}(V)$ and $g^\dagger(V)$ lie within the $\ep$-Wasserstein ball around $X_T^\delta$. 

The cost of $X_w$ is $\E[\varsigma_T X_w] = w\;\E[\varsigma_T \tilde{g}(V)] + (1 - w) \;\E[\varsigma_T X^\dagger]$. Thus, the cost of $X_w$ is a linear interpolation between $\;\E[\varsigma_T \tilde{g}(V)]$ and $\E[\varsigma_T X^\dagger]$. Moreover, by Lemma \ref{lemma-optimal-budget} it holds that $\E[\varsigma_T \tilde{g}(V)] < \E[\varsigma_T \tilde{X}] \le X_0^\delta$. Thus, if $\E[\varsigma_T X^\dagger] \le X_0^\delta$, then $X^\dagger$ fulfils all constraints of optimisation problem \eqref{eqn: main-formulation-no-buget} and has a strictly smaller risk measures than $\tilde{X}$, providing a contradiction. If $\E[\varsigma_T X^\dagger] > X_0^\delta$, then there exists a $w^*$ such that $\E[\varsigma_T X_{w^*}] = X_0^\delta$. Thus, the random variable $X_{w^*}$ satisfies all constraints of optimisation problem \eqref{eqn: main-formulation} and, moreover, it has a strictly smaller risk measure than $\tilde{X}$ as
\begin{equation}
    \RM[X_{w^*}]
    = w^*\; \RM[\tilde{g}(V)] + (1 - w^*) \;\RM[ g^\dagger(V)]
    < w^*\; \RM[\tilde{X}] + (1 - w^*)\; \RM[\tilde{X}]
    =  \RM[\tilde{X}]\,.
\end{equation}
This is a contradiction to the optimality of $\tilde{X}$ and thus concludes the proof.
\Halmos

\subsubsection*{Proof for Proposition \ref{prop:quantile-rep}}
This Proposition is a combination of well-known results, see e.g. Chapter 1 of \cite{Ruschendorf2013Springer}. We provide a proof, as a proof of the exact statement of Proposition \ref{prop:quantile-rep} is not easily available. Recall that $\Coninv(\cdot\,|\,\cdot)$ is a conditional quantile function and that, e.g. Theorem 1.12 in \cite{Ruschendorf2013Springer}, the random variable $\Coninv\left(U\, |\,  U^\delta\right)$ is $\P$-independent of $U^\delta$ and follows a standard uniform distribution under $\P$. Therefore, $X \stackrel{\P}{\sim} F_X$. Next, we show that $(X, X_T^\delta)$ has copula $\Co$. For this we calculate, using independence of $U$ and $U^\delta$,
\begin{align}
    \P\left(X \le x, \, X_T^\delta \le y\right)
    &= \P\left(\Coninv \left(U\,|\, U^\delta\right) \le F_X(x), \, U^\delta \le F(y)\right)\\
    &= \int_0^1 \int_0^1 \Id_{\left\{\Coninv \left(u\,|\, v\right) \le F_X(x)\right\}} \, \Id_{\{v \le F(y)\}}\, du \, dv\\
    &= \int_0^1 \int_0^1 \Id_{\left\{u \le \Con \left(F_X(x)\,|\, v\right) \right\}} \, du\, \Id_{\{v \le F(y)\}} \, dv\\
    &= \int_0^{F(y)}  \Con \left(F_X(x)\,|\, v\right)\, dv\\
    &= \Co \left(F_X(x)\,, F(y)\right)\,,
\end{align}
where the last equality follows by the definition of conditional distribution functions. 
 
Next we show that any random variable $Y \stackrel{\P}{\sim}F_X$ where $(Y,X_T^\delta)$ has copula $\Co$ has representation $Y = \Finv_X\left(\Coninv\left(U\, |\,  U^\delta\right)\right)$. Applying Theorem 1.10 in \cite{Ruschendorf2013Springer}, it holds that for any $U \stackrel{\P}{\sim} \U(0,1)$ independent of $U^\delta$
\begin{equation}
    \left(X_T^\delta, \, Y\right)
    \stackrel{\P}{=} 
    \left( X_T^\delta, \, \Finv_{X | X_T^\delta}(U\,|\,X_T^\delta )\right) \,,
\end{equation} 
where $\stackrel{\P}{=} $ means equal in distribution under $\P$. By definition of the conditional copula, 
\begin{equation}
    F_{X | X_T^\delta}(x\,|\,y )
    = \Con \left(F_X(x)\,|\, F(y)\right)
\end{equation}
and taking the inverse in the first argument, we obtain 
\begin{equation}
    \Finv_{X | X_T^\delta}(u\,|\,y )
    = \Finv_X\left(\Coninv\big(u\, |\, F(y)\big)\right)\,.
\end{equation}
And thus $\Finv_{X | X_T^\delta}(U\,|\,X_T^\delta )  = \Finv_X\left(\Coninv\left(U\, |\,  U^\delta\right)\right)$ $\P$-a.s. The representation of $\C^\delta$ is immediate.
\Halmos

\begin{lemma}[Improved Budget Constraint]\label{lemma-optimal-budget}
For any random variable $X \stackrel{\P}{\sim} F_X$ with $X \in \C^\delta$ it holds that
\begin{equation}
        \E[\, \varsigma_T X\,] \ge     \E[\, \varsigma_T \tilde{X}\,]\,,
\end{equation}
where $\tilde{X} = \Finv_X(V)$ and equality holds if and only if $X \in \tilde{\C}^\delta$ which implies that $\tilde{X} = X\;\P$-a.s..
\end{lemma}

\subsection*{Proof of Lemma \ref{lemma-optimal-budget}}
First note that $\tilde{X} = \Finv_X(V)$ is $\F_T$-measurable and $\tilde{X}\subset \tilde{\C}^\delta$. Moreover, by Proposition \ref{prop:quantile-rep} there exists a $U$ independent of $U^\delta$ such that $X = \Finv_X\left(\Coninv \left(U \,|\,U^\delta\right)\right)\; \P$-a.s.. Thus we can write
\begin{align}
    \E[\, \varsigma_T X\,] 
    &= \E\left[\, \varsigma_T \,\Finv_X\left(\Coninv (U \,|\,U^\delta)\right)\,\right]
    = \E\left[\, \E\left[\,\varsigma_T\, \Finv_X\left(\Coninv (U \,|\,U^\delta)\right)\right]\,\left|\, U^\delta\,\right]\,\right]\,.
\end{align}    
Next, recall that for any two random variables $Y_1, Y_2$ it holds \citep{Ruschendorf1983Metrika}
\begin{equation}\label{eq:lower-mean}
    \E\left[Y_1\,Y_2\right] 
    \ge \E\left[Y_1^\prime\,Y_2^\prime\right]\,,
\end{equation}
 where $Y_1^\prime \stackrel{\P}{=} Y_1$, $Y_2^\prime \stackrel{\P}{=} Y_2$ and the vector $(Y_1^\prime \,, Y_2^\prime)$ is counter-monotonic. The vector $(Y_1^\prime \,, Y_2^\prime)$ is counter-monotonic, if $Y_2^\prime = h(Y_2^\prime)$, with $h$ decreasing. Moreover, Equation \eqref{eq:lower-mean} holds with equality if and only if $(Y_1\,,Y_2)$ is counter-monotonic.

Next we show that the two random variables
\begin{equation}
   \varsigma_T\,|_{ U^\delta}
    \quad \text{and} \quad
     \Finv_X\left(\left.\Coninv \left(\left.1 - F_{\varsigma_T | U^\delta}(\varsigma_T\,|\,U^\delta) \,\right|\,U^\delta\right)\right)\, \right|_{U^\delta }
\end{equation}
are counter-monotonic. Indeed, note that the distribution functions $\Finv_X$ and $F_{\varsigma_T | U^\delta}$ are increasing, and that $\Coninv $ is increasing in its first argument. Thus, the composition $\Finv_X \circ \Coninv \circ (1 - F_{\varsigma_T | U^\delta}(\cdot))$ is decreasing and therefore the two random variables are counter-monotonic. Combining the above arguments, we obtain that
\begin{align}    
    \E[\, \varsigma_T X\,] 
    &= \E\left[\, \E\left[\,\varsigma_T\, \Finv_X\left(\Coninv (U \,|\,U^\delta)\right)\right]\,\left|\, U^\delta\,\right]\,\right]\\
    & \ge \E\left[\, \E\left[\,\varsigma_T\, \Finv_X\left(\left.\Coninv \left(1 - F_{\varsigma_T | U^\delta}(\varsigma_T\,|\,U^\delta) \,\right|\,U^\delta\right)\right)\,\left|\, U^\delta\,\right]\right]\right.\\
    &= \E\left[\, \varsigma_T \,\Finv_X\left( V \right)\,\right]\\[0.5em]
    &= \E[\, \varsigma_T \tilde{X}\,] \,,
\end{align}
Moreover, equality holds if and only if $U =1 - F_{\varsigma_T | U^\delta}(\varsigma_T\,|\,U^\delta)$, which implies that $X = \tilde{X}\; \P$-a.s. and $X \in \tilde{C}^\delta$. 
\Halmos


\subsection*{Proof of Proposition \ref{prop:c-tilde-delta}}
Let $X\stackrel{\P}{\sim}F_X$ fulfil the constraints of Problem \eqref{eqn: main-formulation-Pprime} and define the family of random variable $X_c := \Finv_X (V) + c$, for $c\ge 0$. We assume w.l.o.g. that $X \in \C^\delta / \tilde{\C}^\delta$ and split the proof into two parts; the first part considers $X$ in the interior of the Wasserstein ball and the second part we study $X$ at the boundary of the Wasserstein ball.

\underline{Part I: Interior of Wasserstein ball:}
First, we consider the case where $d_2\left[\, X_T^\delta\,,\,X\right] = \ep^\prime < \ep$, i.e., $X\in\oM_{\ep'}$. Since for any   $c \ge 0$, it holds that $\Finv_{X_c}(u) = \Finv_X(u) + c$, we obtain that $X_c \in \tilde{\C}^\delta\subset \C^\delta$. By Lemma \ref{lemma-optimal-budget} it holds that
\begin{subequations}
\begin{align}
     \E\, [\varsigma_T \, \Finv_X (V)\,] 
    \;< \;\E\, [\varsigma_T \, X\,] 
    \;\le \;X_0^\delta,
\end{align}
and 
\begin{align}
    \E\, [\varsigma_T \, X_c\,]
    = \E\, [\varsigma_T \, \Finv_X (V)\,] + c\,.
\end{align}%
\label{eqn:budget-bound}
\end{subequations}%
The Wasserstein distance between $X_T^\delta$ and $X_c$ is 
\begin{align}
    \left(d_2\left[X_T^\delta\,, \, X_c\right]\right)^2
    &= \int_0^1 \left(\Finv(u) -\Finv_{X}(u) - c\right)^2\, du\\
    &= \left(d_2\left[X_T^\delta\,, \, X\right]\right)^2 
        - 2\,c\, \int_0^1 \left( \Finv(u) -\Finv_{X_c}(u) \right)\, du   + c^2\,.
\end{align}
By Proposition \ref{prop: range-P-mean}, the maximal absolute distance between the mean of $X_T^\delta$ and a random variable with quantile function in $\oM_{\ep^\prime}$ is equal to $\ep^\prime$. Therefore, we obtain the bound
\begin{equation}
\label{eqn:Wasser-bound}
    \left(d_2\left[X_T^\delta\,, \, X_c\right]\right)^2
    \le (\ep^\prime)^2 
    + 2\,c \,\ep^\prime + c^2
     = \left(\ep^\prime + c\right)^2\,.
\end{equation}
Thus, by \eqref{eqn:budget-bound} and \eqref{eqn:Wasser-bound}, the choice $c := \min\left\{ \,X_0^\delta - \E\, [\varsigma_T \, \Finv_X (V)\,] \,,\, \ep - \ep^\prime\,\right\} >0$ implies that $X_c$ satisfies all the constraints of \eqref{eqn: main-formulation-Pprime} and $\RM[X_c]  = \RM[X] - c < \RM[X]\,$.

\underline{Part II: Boundary of Wasserstein ball:} If  $d_2\left[\, X_T^\delta\,,\,X\right] = \ep$, then $X_0 =\Finv_X(V)$ satisfies all the constraints (in particular has a strictly smaller cost than $X$) and $\RM[X_0] = \RM[X]$.
\Halmos

\begin{lemma}[Improving the Benchmark's Risk Measure]\label{lemma:risk-measure}
The optimal random variable $X_T^*$ attaining \eqref{eqn: main-formulation-Pprime} satisfies $\RM[X_T^*] \le \RM[X_T^\delta]$.
\end{lemma}
\subsubsection*{Proof of Lemma \ref{lemma:risk-measure}}
We construct an admissible random variable that has risk measure at most that of the benchmark's terminal wealth. For this, consider the random variable $\Finv(V)$ with quantile function $\Finv$. Clearly, $\Finv \in \oM_\ep$, $\Finv(V)\in \tilde{\C}^\delta$, and by Lemma \ref{lemma-optimal-budget}
\begin{equation}
    \E[\varsigma_T \,\Finv(V)] \le \E[\varsigma_T \,\Finv(U^\delta)] = \E[\varsigma_T \,X_T^\delta ] = X_0^\delta\,,
\end{equation}
thus $\Finv(V)$ fulfils all the constraints of \eqref{eqn: main-formulation-Pprime}. Moreover, as $\Finv(V) \stackrel{\P}{=} X_T^\delta$, we obtain $\RM[X_T^\delta] = \RM[\Finv(V)]$.
\Halmos

\subsubsection*{Proof of Theorem \ref{thm:quantile-formulation}}
The proof proceeds by first showing that the sets $\oM^\delta$ and 
\begin{equation} 
     \B^\delta := \tilde{\C}^\delta  \cap \{\,X\in \Lp(\P,\Omega) \,|\,\E[\,\varsigma_T\,X\,]\le X_0^\delta \,\}
\label{eqn:C-delta-cap-FT-cap-budget}
\end{equation}
are isomorpohic. That is, there is a one-to-one and onto mapping between (a) the set of square integrable $\F_T$-measurable random variables satisfying the budget constraint and that have copula $\Co$ with $X_T^\delta$, and (b) the set of quantile functions satisfying the constraint in the definition of $\oM^\delta$.

We construct the isomorphism, which we denote $\varphi:\B^\delta\to \oM^\delta$ explicitly as follows. Take any  $X\in\B^\delta$, so that $(X,X_T^\delta)$ has copula $\Co$. Therefore  we have $X=\Finv_X(V)$ a.s., and the expectation in the budget constraint may be written as
\begin{align}
\E[\, \varsigma_T\,X\,] 
&=  
\E\left[\,\varsigma_T\, \Finv_X(V)\,\right]
=
\E\left[\,\Finv_X(V)\;\E\left[\left. \,\varsigma_T\,\right|\, \sigma(V)\right]\right]\,,
\label{eqn:EX-in-terms-of-cond-RN-midstep}
\end{align}
where in the last equality, $\sigma(V)$ denotes the natural $\sigma$-algebra generated by $V$. To proceed, we claim that
\begin{equation}
\label{eqn:conditional-dq-dp}
\E\left[\left. \,\varsigma_T\,\right|\, \sigma(V)\right]
= \hat{f}_V(V),
\end{equation}
with $\hat{f}_V(u):=\frac{d}{du}\mathbb{E}\left[\varsigma_T\Id_{\{V \le u\}}\right]$. To prove equality \eqref{eqn:conditional-dq-dp}, it suffices to show  the lhs and rhs are equal under expectation when multiplied by $\psi(V)$, where $\psi:\R\to\R$ is any Borel-measurable function. For the rhs we first note that 
\begin{subequations}
\begin{align}
    \hat{f}_V(u) &= \frac{d}{du}\E[\,\varsigma_T\,\Id_{\{V \le u\}}\,]\\
        &= \frac{d}{du}\E\left[\,\frac{\varsigma_T}{\E[\varsigma_T]}\,\Id_{\{V \le u\}}\,\right]\,\E[\varsigma_T]\\
        &= \frac{d}{du}\E^{\Q_T}\left[\,\Id_{\{V \le u\}}\,\right]\,\E[\varsigma_T]\\
        &= f^{\Q_T}_V(u)\,\E[\varsigma_T]\,,
\end{align}%
\end{subequations}%
where $\Q_T$ denotes the probability measure induced by the Radon-Nikodym derivative $\frac{d\Q_T}{d\P} = \frac{\varsigma_T}{\E[\varsigma_T]}$ and $f^{\Q_T}_V$ is the $\Q_T$-density of $V$. Thus, we obtain
\begin{subequations}
\begin{align}
 \E\left[\,\hat{f}_V(V)\, \psi(V)\,\right]
    &= \int_0^1  f^{\Q_T}_V(u)\,\psi(u)\,du\;\E[\varsigma_T]\\
    &=\E^{\Q_T} [\psi(V) ]\;\E[\varsigma_T] \\
    &=\E \left[\frac{\varsigma_T}{\E[\varsigma_T]}\psi(V) \right]\;\E[\varsigma_T] \\
    &= \E[\,\varsigma_T\,\psi(V)\,]\,.
    \label{eqn:cond-dqdp-pt1}
\end{align}%
\end{subequations}%
For the lhs we have
\begin{subequations}
\begin{align}
    \E\left[\,\E\left[\,\left.\varsigma_T \,\right|\, \sigma(V)\,\right] \psi(V)\,\right] 
   =
    \E\left[\,\E\left[\,\left.\varsigma_T \, \psi(V) \,\right|\, \sigma(V)\,\right]\,\right] 
= \E\left[\,\varsigma_T \, \psi(V) \,\right].
    \label{eqn:cond-dqdp-pt2}
\end{align}
\end{subequations}
As both expression hold for arbitrary $\psi$, equality \eqref{eqn:conditional-dq-dp} holds.

Returning to \eqref{eqn:EX-in-terms-of-cond-RN-midstep}, we  have
\begin{equation}
    \E[\,\varsigma_T\,X\,] = \E \left[\,\Finv_X( V)\,\hat{f}_V(V)\,\right]
    = \int_0^1 \Finv_X(u)\,\xi(u)\,du.
    \label{eqn:EQX-equals-int-Finv-xi}
\end{equation}
Hence, for all $X\in\B^\delta$ we have $\E[\,\varsigma_T\,X\,]=\int_0^1 \Finv_X(u)\,\xi(u)\,du\le X_0^\delta$. Next, define the function $\varphi(X):=\Finv_X$, which is non-decreasing. As $X\in\L^2(\Omega,\P)$, $\varphi(X)\in\Lp([0,1])$. Putting this together with the budget constraint equivalence, we have $\varphi(X)\in\oM^\delta$.

We can construct the inverse map $\varphi^{-1}:\oM^\delta\to\B^\delta$ as follows. Take any $g\in\oM^\delta$, then define $\varphi^{-1}(g):=g(V)$. As $V$ is $\F_T$-measurable, so is $\varphi^{-1}(g)$. Moreover $g\in \Lp([0,1])$, thus $\varphi^{-1}(g)\in\Lp(\Omega, \P)$. As $\int_0^1g(u)\xi(u) \le X_0^\delta$ we have (using \eqref{eqn:EQX-equals-int-Finv-xi} in reverse)  $\E[\,\varsigma_T\,\varphi^{-1}(g)\,]\le X_0^\delta$. Finally, as $g$ is increasing, the random vector $(\varphi^{-1}(g),\,X_T^\delta)=(g(V),\,X_T^\delta)$ has the same copula as $(V,\,X_T^\delta)$, which indeed has copula $\Co$. Consequently, $\varphi^{-1}(g)\in\B^\delta$.

The set $\B^\delta$ contains all random variables satisfying the constraints in \eqref{eqn: main-formulation-Pdagger} except for the Wasserstein distance constraint. Moreover, the Wasserstein distance and the risk measure only depend on quantile functions (and not random variables). Thus, we have established the equivalence of \eqref{eqn: main-formulation-Pprime} and \eqref{eqn: main-formulation-Pdagger}, as detailed in the statement.
\Halmos

\subsubsection*{Proof of Theorem \ref{thm:Existence-Uniqueness-P1p}.}
First, as the benchmark $\Finv\in\oM_\ep\cap\oM^\delta$ the constraint set is non-empty. The remainder of the proof is broken into two parts.
\begin{enumerate}
    \item[I.] \textit{Convexity of constraint set}. 
    Take $g_1,g_2\in\oM_\ep\cap\oM^\delta$ and define the convex combination $g^\omega:=\omega\,g_1 + (1-\omega)\,g_2$ for $\omega\in[0,1]$. Clearly, $g^\omega \in \oM^\delta$.
    Furthermore, 
    \begin{align*}
    \int_0^1  \big( \Finv(u)& -g^\omega(u)\big)^2\,du 
    \\
    \begin{split}
    =& \; \omega^2\,\int_0^1 \left(\Finv(u) -g_1(u)\right)^2\,du
        +
    (1-\omega)^2\,\int_0^1 \left(\Finv(u) -g_2(u)\right)^2\,du
    \\
    &  + 2\, \omega\, (1-\omega) \int_0^1 \left(\Finv(u) -g_1(u)\right)\left(\Finv(u) -g_2(u)\right)\, du
    \end{split}
    \\
    \text{\tiny(as $g_1,g_2\in\oM_\ep$) }\le&  \, \omega^2 \ep^2  + (1-\omega)^2 \ep^2 + 2\, \omega\, (1-\omega) \int_0^1 \left(\Finv(u) -g_1(u)\right)\left(\Finv(u) -g_2(u)\right)\, du
    \\
    \begin{split}
    \text{\tiny(by Young's ineq.) }\le&  \, \ep^2  \, ( \omega^2 + (1- \omega)^2)
    \\ & \quad
    +  \omega\, (1-\omega)\int_0^1  \left(\left(\Finv(u) -g_1(u)\right)^2   
    + \left(\Finv(u) -g_2(u)\right)^2\right)du
    \end{split}
    \\    
    \le&  \, \ep^2 \, ( \omega^2 + (1- \omega)^2) +  2\, \omega\, (1-\omega) \,\ep^2
    \\
    =&  \, \ep^2.
    \end{align*}
    Therefore, $g^{\omega}\in\oM_\ep$, hence $g^{\omega}\in\oM_\ep \cap \oM^\delta$ and $\oM_\ep \cap \oM^\delta$ is a convex set.
    
    \item[II.] \textit{Existence \& Uniqueness} 
        As $\oM_\ep \cap \oM^\delta$ is convex and, due to the regularisation term, the objective function 
        is strictly convex in $g$,  the solution to \eqref{eqn: main-formulation-Pdagger-c} exists and is unique. 
\end{enumerate}
\Halmos

\subsubsection*{Proof of Theorem \ref{thm:isotonic-Pdagger-c}.}
The proof is split in two parts. First, we show that for any $c$ the quantile function is of the form \eqref{eqn: pf-gstar}. In a second step, we prove the existence of $\bar{c}$ such that for all $c\in(0,\bar{c}]$, the Wasserstein constraints is binding, i.e., $\lambda_1>0$. 

\underline{Part I:}
Let $c>0$, then the objective functional pertaining to \eqref{eqn: main-formulation-Pdagger-c} with Lagrange multipliers $\lambda_1,\lambda_2\ge 0$ is
\begin{equation}
\begin{split}
    J(g\,, \lambda_1,\,  \lambda_2):=&
    - \int_0^1 g(u) \gamma(u) \, du + \left(\lambda_1 + c\right) \left( \int_0^1 \left(g(u) - \Finv(u)\right)^2 \, du  - \ep^2 \right)
    \\
    &+ 
    \lambda_2 \left( \int_0^1 g(u) \, \xi(u)\,du - X_0^\delta \right) + c \,\ep^2.
\end{split}    
\end{equation}
Next, completing the square and collecting terms, yields
\begin{subequations}
\begin{align}
J(g\,, \lambda_1,\,  \lambda_2)=&\, 
    (\lambda_1+c) \int_0^1 \left(g(u) - \left( \Finv(u) + \tfrac{1}{2(\lambda_1 + c)}\left(\gamma(u) -\lambda_2\,\xi(u) \right)\right) \right)^2 \, du
    \label{eqn:P1p-proof-line1}
    \\
    &\,
    - (\lambda_1+c)\int_0^1 \left( \Finv(u) + \tfrac{1}{2(\lambda_1+c)}\left(\gamma(u) -\lambda_2\,\xi(u) \right)\right)^2\, du 
    \\
    &\,
    + (\lambda_1 + c)\int_0^1\left(\Finv(u)^2 - \ep^2 \right) \, du- \lambda_2 \, X_0^\delta\,+ c \,\ep^2 .
\end{align}
\end{subequations}
Hence, \eqref{eqn: main-formulation-Pdagger-c} can be written as
\begin{equation}
    \sup_{\lambda_1, \lambda_2  \ge 0}\,  \inf_{g \in \oM} J(g, \, \lambda_1\, , \lambda_2)\,.
\end{equation}
For fixed $\lambda_1, \lambda_2$, the inner optimisation is attained by optimising \eqref{eqn:P1p-proof-line1}, as all other terms are independent of $g$. Moreover, the inner optimisation is attained by the isotonic projection of $\ell(u;\lambda_1,\lambda_2,c)$ onto $\oM$ (see Definition \ref{def:isotonic}), resulting in \eqref{eqn: pf-gstar}. 
By Theorem \ref{thm:Existence-Uniqueness-P1p}, this optimum is unique.

\underline{Part II:} Next, we show that for small enough $c$, the Lagrange multiplier $\lambda_1$ must be strictly positive. Throughout the proof, we write $\ell_c(u):  = \ell(u;\lambda_1,\lambda_2,c)$ and $\ell^\uparrow_c(u):  = \ell^\uparrow(u;\lambda_1,\lambda_2,c)$. For this, assume that $\lambda_1 = 0$, then
\begin{equation}
    \ell_c(u):= \Finv(u) + \tfrac{1}{2 c}\left(\gamma(u) -\lambda_2 \xi(u) \right)
    = \frac{1}{2\,c}k_c(u),
\end{equation}
where $k_c(u):=2\,\,c\,\Finv(u) + \gamma(u) -\lambda_2 \xi(u) $. Note that $\ell^\uparrow_c(u) =\frac{1}{2\,c} k^\uparrow(u)$. Next, the Wasserstein distance between $\ell^\uparrow_c$ and $\Finv$ may be written as
\begin{equation}
    d_2(\ell^\uparrow_c,\,\Finv) = \frac{1}{4\,c^2}\int_0^1\left(k_c^\uparrow(u)-2\,c\,\Finv(u)\right)^2\,du\,.
\end{equation}
Now define $A_c:=\int_0^1(k_c^\uparrow(u)-2\,c\,\Finv(u))^2\,du$, due to the integrability assumptions, we have that
\begin{equation}\label{pf:eq:lambda1>0}
    \lim_{c\searrow 0}A_c
    =
    \int_0^1    \lim_{c\searrow 0}\left(k_c^\uparrow(u)-2\,c\,\Finv(u)\right)^2\,du
     =
    \int_0^1\left(k^\uparrow_0(u)\right)^2\,du >0\,,
\end{equation}
where the strict inequality follows from $k^\uparrow_0(u) = \left(\gamma(u) - \lambda_2 \xi(u)\right)^\uparrow$ and the following argument. If $\lambda_2 = 0$, then the strict inequality is immediate. Next, consider the case that $\lambda_2>0$ and $\gamma(u) = \lambda_2 \xi(u)$, so that $k^\uparrow_0(u)=0$ and $\ell^\uparrow_0 (u)= \Finv(u)$. If further 
\begin{enumerate}[label = $(\roman*)$]
    \item $\Co $ is the comonotonic copula, then $\lambda_2 = (\int_0^1 \xi(v)\, dv)^{-1}$ (since $\gamma$ integrates to 1), which is a contradiction to Assumption \ref{asm:como-gamma}.
    
    \item $\Co$ is not the comonotonic copula, then the cost of the optimal strategy is $\int_0^1 \ell_0^\uparrow(u) \xi(u) \, du = \int_0^1 \Finv(u) \xi(u) \, du < X_0^\delta$ by Lemma \ref{lemma-optimal-budget}. This however, is a contradiction to $\lambda_2 >0$, which requires that the budget constraint is binding. 
\end{enumerate}

Therefore \eqref{pf:eq:lambda1>0} holds and since $d_2(\ell^\uparrow_c,\Finv)=\frac{1}{4c^2}A_c$, there must exist a $\bar{c}>0$ such that for all $c \in (0, \bar{c})$ it holds that  $d_2(\ell^\uparrow_c,\Finv)>\ep^2$, and we obtain a violation of the Wasserstein constraint and that $\lambda_1 = 0$.
\Halmos

\subsubsection*{Proof of Corollary \ref{cor: solution-to-Pdagger-c}}
For $c \in (0, \bar{c})$, from Theorem \ref{thm:isotonic-Pdagger-c}, the Wasserstein constraints in \eqref{eqn: main-formulation-Pdagger-c} is binding. Therefore, the optimisation \eqref{eqn: main-formulation-Pdagger-c} reduces to 
\begin{equation}
\label{eqn:Pdagger-c-binded}
\left(\inf_{g\in\oM_\ep\cap \oM^\delta}
 \RM(g) \right) + c \, \varepsilon^2.
\end{equation}
The above shows that the objective functions (and hence the infimum) of \eqref{eqn: main-formulation-Pdagger-c} and \eqref{eqn: main-formulation-Pdagger} are equal up to a constant, consequently, a solution to   \eqref{eqn: main-formulation-Pdagger-c} is indeed a solution to \eqref{eqn: main-formulation-Pdagger} and vice versa. From \eqref{eqn:Pdagger-c-binded} the infimum in \eqref{eqn: main-formulation-Pdagger-c} is independent of $c$.  Further, Theorem \ref{thm:Existence-Uniqueness-P1p} shows that the solution to  \eqref{eqn: main-formulation-Pdagger-c} is unique. As the solution to \eqref{eqn: main-formulation-Pdagger-c} and \eqref{eqn: main-formulation-Pdagger} are equivalent, we established uniqueness of \eqref{eqn: main-formulation-Pdagger}. Applying Theorem \ref{thm:quantile-formulation}, we obtain the unique solution to \eqref{eqn: main-formulation-Pdagger}.
\Halmos

\subsubsection*{Proof of Proposition \ref{prop:c-tilde-delta-all}.}
The proof follows along the line of the proof of Proposition \ref{prop:c-tilde-delta}. 

Let $X\stackrel{\P}{\sim}F_X$ fulfil the constraint of Problem \eqref{eqn: main-formulation-Pall} and define the family of random variables $X_c := \Finv_X (V) + c$, for $c\ge 0$ and $V$ as defined in Proposition \ref{prop:c-tilde-delta-all}, Equation \eqref{eq:c-tilde-delta-all}. We assume w.l.o.g. that $X \neq X_c$ $\P$-a.s. for all $c \ge 0$. The proof is split into two parts; first we considers $X$ in the interior of the Wasserstein ball and second we study $X$ at the boundary of the Wasserstein ball.

\underline{Part I: Interior of Wasserstein ball:}
Consider the case where $d_2\left[\, X_T^\delta\,,\,X\right] = \ep^\prime < \ep$. Since for any   $c \ge 0$, it holds that $\Finv_{X_c}(u) = \Finv_X(u) + c$, we obtain that $X_c \in \tilde{\C}^\delta $, where $\tilde{\C}^\delta $ is defined in Proposition \ref{prop:c-tilde-delta-all}, Equation \eqref{eq:c-tilde-delta-all}. Then, since counter-monotonic random variables have the smallest expectation, see Equation \eqref{eq:lower-mean}, it holds that
\begin{align}
     \E\, [\varsigma_T \, \Finv_X (V)\,] 
    \;= \;\E\, [\varsigma_T \, \Finv_X(1 - F_{\varsigma_T}(\varsigma_T))\,] 
    \;< \;\E\, [\varsigma_T \, X\,] 
    \;\le \;X_0^\delta\,.
\end{align}
The remainder of the proof, including part II, is identical to the proof of Proposition \ref{prop:c-tilde-delta}.
\Halmos

\subsubsection*{Proof of Proposition \ref{prop: solution-all}.}
This follows from the observation that Theorems \ref{thm:quantile-formulation}, \ref{thm:Existence-Uniqueness-P1p}, and \ref{thm:isotonic-Pdagger-c}, as well as Corollary \ref{cor: solution-to-Pdagger-c}, only depend on $V$ through the definition of $\xi(u)$ in Equation \eqref{eqn:conditional-R-N}.
\Halmos

\subsubsection*{Proof of Proposition \ref{prop: range-P-mean}.}
We first solve the optimisation for $\ml$. Consider the objective functional
\begin{align}
J(g\, , \zeta) :&= \int_0^1 g(u)\, du + \zeta \left( \int_0^1 \left(g(u) - \Finv(u)\right)^2 \, du  - \ep^2 \right)
\\
&= \zeta \int_0^1 \left( g(u) -  \left( \Finv(u) - \tfrac{1}{2\, \zeta} \right) \right)^2\, du  
+ \int_0^1 \Finv(u)\, du - \tfrac{1}{4\, \zeta}  - \zeta\,  \ep^2\,. 
\end{align}
The parameter $\zeta$ plays the role of a Lagrange multiplier, and we have that 
\begin{equation}\label{eqn: proof-isotonic-Euler}
\inf_{g\in\oM_\ep}\, \int_0^1 g(u)\,du
= 
\sup_{\zeta \ge 0} \inf_{g \in \oM}J(g, \, \zeta) 
\end{equation}
Note that, for $\zeta\ge 0$ fixed, 
\begin{equation}
    \argmin_{g\in\oM} J(g,\zeta) = \argmin_{g\in\oM} \int_0^1 \left( g(u) -  \left( \Finv(u) - \tfrac{1}{2\, \zeta} \right) \right)^2\, du
\end{equation}
and the rhs is attained by the isotonic projection of $\Finv - \tfrac{1}{2\, \zeta}$ onto the non-decreasing functions. Since $\Finv$ is non-decreasing, the isotonic projection is precisely $g_{\zeta}(u):=\Finv(u) - \tfrac{1}{2\, \zeta}$. Inserting this optimum into the objective function and optimising over $\zeta$, we have
\begin{equation}
\ep = \left(\int_0^1 \left( g_{\zeta}(u) - \Finv(u)\right)^2 \, du\right)^\frac12
    = \tfrac{1}{2\, \zeta}.
\end{equation}
Therefore $\underline{g}(u) = \Finv(u)  - \ep$ and $\ml = \m - \ep$. The solution for $\mr$ follows analogously.
\Halmos

\subsubsection*{Proof of Proposition \ref{prop: range-P-sd}.}
First, we solve the optimisation for $\sl^2$. The corresponding objective functional is then
\begin{align}
J(g\, , \zeta_1\, , \zeta_2) :=& \int_0^1 \left(g(u) - m\right)^2\, du + \zeta_1  \left( \int_0^1 \left(g(u) - \Finv(u)\right)^2 \, du  - \ep^2 \right)
+ \zeta_2 \left( \int_0^1 g(u)\, du - m\right)
\\
=\, & (\zeta_1  + 1)\int_0^1 \left( g(u) - \tfrac{1}{\zeta_1  + 1}\left(\zeta_1\Finv(u) - \tfrac{\zeta_2}{2} + m \right) \right)^2\, du  
- \tfrac{1}{\zeta  + 1}\int_0^1\left( \zeta_1\Finv(u) - \tfrac{\zeta_2}{2} + m \right)^2\, du 
\\
&+ \zeta_1 \int_0^1 \left(\Finv(u)^2 - \ep^2\right)\, du + m^2  - \zeta_2 m\,. 
\end{align}
The parameters $\zeta_1, \zeta_2$ play the role of a Lagrange multiplier, and we have that, for $\zeta_1\ge0 $ and $\zeta_2\in \R$ fixed, 
\begin{equation}
    \argmin_{g\in\oM_\ep} J(g\, ,\zeta_1\, , \zeta_2) = \argmin_{g\in\oM_\ep} \int_0^1 \left( g(u) -  \tfrac{1}{\zeta_1  + 1}\left(\zeta_1\Finv(u) - \tfrac{\zeta_2}{2} + m \right) \right)^2\, du.
\end{equation}
The rhs is attained by the isotonic projection of $g_{\zeta_1, \zeta_2}(u):= \tfrac{1}{\zeta_1  + 1}\left(\zeta_1\Finv(u) - \tfrac{\zeta_2}{2} + m \right) $, which is, by increasingness of $\Finv$, equal to $g_{\zeta_1, \zeta_2}(u)$. Enforcing the mean and Wasserstein constraints and simplifying, we obtain
\begin{equation}
\frac{\zeta_2}{2} = \zeta_1 \Delta m\, , 
\quad \text{and}\quad
\left(\zeta_1 + 1\right)^2 \left(\ep^2 - (\Delta m)^2\right) = \s\, .
\end{equation}

If $\s \ge \sqrt{ \, \ep^2  - (\Delta m)^2\, }$, then $\zeta_1$ is the positive root, given by 
\begin{equation}
    \zeta_1  = \frac{\s}{\sqrt{ \, \ep^2  - (\Delta m)^2\, }} - 1.
\end{equation}
Direct calculations provide $\underline{s}>0$ in \eqref{prop: range-P-sd-sd-values} and $\underline{g}$ in \eqref{prop: range-P-sd-gl}. 

If $\s < \sqrt{ \, \ep^2  - (\Delta m)^2\, }$, then $\zeta_1 = 0$ and by enforcing the mean constraint, $\zeta_2 = 0$ and thus, $\underline{g}(u) = m$ and $\underline{s} = 0$.

Next, we solve the optimisation for $\overline{s}^2$, which is similar to the proof of $\underline{s}$. The objective function is
\begin{align}
J(g\, , \zeta_1\, , \zeta_2) :=& -\int_0^1 \left(g(u) - m\right)^2\, du + \zeta_1  \left( \int_0^1 \left(g(u) - \Finv(u)\right)^2 \, du  - \ep^2 \right)
+ \zeta_2 \left( \int_0^1 g(u)\, du - m\right)
\\
=\, & (\zeta_1  - 1)\int_0^1 \left( g(u) - \tfrac{1}{\zeta_1  - 1}\left(\zeta_1\Finv(u) - \tfrac{\zeta_2}{2} - m \right) \right)^2\, du  
- \tfrac{1}{\zeta  - 1}\int_0^1\left( \zeta_1\Finv(u) - \tfrac{\zeta_2}{2} - m \right)^2\, du 
\\
&+ \zeta_1 \int_0^1 \left(\Finv(u)^2 - \ep^2\right)\, du - m^2  - \zeta_2 m\,. 
\end{align}
The quantile function in $\oM_\ep$, that attains the supremum of $J(g\, , \zeta_1\, , \zeta_2)$ is the isotonic projection of $g_{\zeta_1, \zeta_2}(u): = \tfrac{1}{\zeta_1  - 1}\left(\zeta_1\Finv(u) - \tfrac{\zeta_2}{2} - m\right)$. As $\zeta_1 \ge 0$ and $\Finv$ is increasing, the isotonic projection of $g_{\zeta_1, \zeta_2}$ is indeed $g_{\zeta_1, \zeta_2}$. 
Enforcing the mean and Wasserstein constraints and simplifying, we obtain
\begin{equation}
\frac{\zeta_2}{2} = \zeta_1 \Delta m\, , 
\quad \text{and}\quad
\left(\zeta_1 - 1\right)^2 \left(\ep^2 - (\Delta m)^2\right) = \s^2\, .
\end{equation}
The roof of $\zeta_1$ that provides the larger standard deviation is 
\begin{equation}
    \zeta_1  = \frac{\s}{\sqrt{ \, \ep^2  - (\Delta m)^2\, }} + 1.
\end{equation}
Direct calculations provide $\overline{s}$ in \eqref{prop: range-P-sd-sd-values} and $\overline{g}$ in \eqref{prop: range-P-sd-gu}.
\Halmos

\subsubsection*{Proof of Corollary \ref{cor: min-max-std-in-oM}.}
Taking the supremum over all $m \in [\ml, \, \mr]$ in \eqref{prop: range-P-sd-sd-values}, we obtain 
\begin{equation}
\sup_{m\in [\ml, \, \mr]}\;\sr = \sup_{m\in [\ml, \, \mr]} \;\s  + \sqrt{\,\ep^2 - (\Delta m)^2 \,} \,
    = \s + \ep\,,
\end{equation}
where the maximum standard deviation is attained for $m = \m$. As $\m \in [\ml, \, \mr]$, the corresponding quantile function $\overline{g}$ defined in \eqref{prop: range-P-sd-gu} is in $\oM_\ep$.

Similarly for the infimum
\begin{equation*}
\inf_{m\in [\ml, \, \mr]}\; \sl =\inf_{m\in [\ml, \, \mr]} \;\max\left\{\s  - \sqrt{\,\ep^2 - (\Delta m)^2 \,} \,,\; 0\, \right\}  
    = \max\left\{\s - \ep \,,\; 0\, \right\} \,,
\end{equation*}
where the minimum standard deviation is attained for $m = \m$ and with corresponding quantile function $\underline{g}\in \oM_\ep$.
\Halmos

\subsection*{Proof of Proposition \ref{prop:GBM-CoIn}}

The GBM market model with constant coefficients assumes
\begin{equation}
    dS_t^i = S_t^i\left(\mu^i\,dt+\sigma^i \,dW_t^i \right),
\end{equation}
and the SDF is 
\begin{equation}
    \varsigma_T:= e^{-rT}\;\frac{d\mathbb Q}{d\mathbb P}=e^{-r\,T}\mathcal{E}\left(-\int_0^T \lambda^\intercal\,dW_s\right)
    =e^{-(r +\frac{1}{2}\lambda^\intercal\rho\lambda )T- \lambda^\intercal W_T},
    \label{eqn:sdf-gbm-const}
\end{equation}
with $\lambda=\rho^{-1}\frac{\mu-r}{\sigma}$ (where, in the ratio, division is meant component-wise). Note that when interest rates are deterministic, then $\frac{d\Q}{d\P}=\frac{d\Q_T}{d\P}$.
From Girsanov's Theorem,
\begin{equation}
    W_t^* := \rho\lambda\, t +W_t
\end{equation}
is a $\mathbb{Q}$-Brownian motion, and we may write
\begin{equation}
    dS_t^i = S_t^i\left(r\,dt+\sigma^i \,dW_t^{*i} \right)\,.
\end{equation}
For a constant benchmark $\delta$, \eqref{eqn:dX_upsilon} may be solved explicitly, and the corresponding terminal wealth is
\begin{equation}
    X_T^\delta = X_0^\delta\;e^{\Gamma \, T + \eta^\intercal\, W_T},
    \label{eqn:bench-gbm-const}
\end{equation}
where $\eta:=(\delta^1\sigma^1,\delta^2\sigma^2,\dots,\delta^d\sigma^d)^\intercal$,
\[
\Gamma := r+\delta^\intercal (\mu-r\mathds{1}) - \tfrac{1}{2} \Psi^2, \qquad\text{and}\qquad\Psi^2 :=\eta^\intercal\rho\eta\,.
\]
Next, we compute the joint distribution function of $\varsigma_T$ and $X_T^\delta$. To this end,
\begin{align}
\mathbb{P}\left(\varsigma_T\le z \,,\, X_T^\delta \le x\right)
&= 
\mathbb{P}\left(-\lambda^\intercal W_T\le
\log(z) + \Lambda\,T \,,\, \eta^\intercal W_T \le \log(\tfrac{x}{X_0^\delta}) - \Gamma\,T\right),
\end{align}
where 
\[
\Lambda:=r+ \tfrac{1}{2}\lambda^\intercal\rho\lambda\,.
\]
Note that as
$
    \begin{pmatrix}
    -\lambda^\intercal W_T
    \\
    \eta^\intercal W_T
    \end{pmatrix}
    \stackrel{\P}{\sim}
    \mathcal{N}\left(
    \begin{pmatrix}
    0
    \\
    0
    \end{pmatrix} 
    ;
\Omega\; T   
    \right),
$
where
$
\Omega:=
    \begin{pmatrix}
    \lambda^\intercal \rho\lambda
    & 
    -\lambda^\intercal \rho\eta
    \\
    -\lambda^\intercal \rho\eta
    &
    \eta^\intercal \rho\eta
    \end{pmatrix}\;T,
$
we have that
\begin{equation}
\mathbb{P}\left(\varsigma_T\le z \,,\, X_T^\delta \le x\right)
= \Phi_{\Omega}\left(\log(z) + \Lambda \, T\;;\;\log(\tfrac{x}{X_0^\delta}) - \Gamma\,T \right),
\end{equation}
where $\Phi_\Omega(\cdot,\cdot)$ is the bivariate cdf of a normal random variable with zero mean and covariance matrix $\Omega$.

By tedious computations, we further have that
\begin{equation}
\partial_x\mathbb{P}(\varsigma_T \le z\;,\;X_T^\delta \le x)
= \frac{\phi\left(
\frac{\log(\tfrac{x}{X_0^\delta})-\Gamma\,T}{\sqrt{\eta^\intercal\rho\eta\,T}}\right)}{x\sqrt{\eta^\intercal\rho\eta\,T}}
\,\Phi\left(
\frac{\log(z)+\Lambda\,T + \frac{\lambda^\intercal\rho\eta}{\eta^\intercal\rho\eta}
\left(\log\tfrac{x}{X_0^\delta}-\Gamma\,T\right)
}{
\sqrt{
\left(
\lambda^\intercal\rho\lambda-\frac{(\lambda^\intercal\rho\eta)^2}{\eta^\intercal\rho\eta}
\right)
T
}
}
\right).
\end{equation}
Hence,
\begin{equation}
\begin{split}
F_{\varsigma_T|U^\delta}(z|u) 
= 
    \mathbb{P}(\varsigma_T \le z\;|\;X_T^\delta=  \Finv_{X_T^\delta}(u))    
=
\Phi\left(
\frac{\log(z)+\Lambda\,T + \frac{\lambda^\intercal\rho\eta}{\eta^\intercal\rho\eta}
\left(\log\tfrac{\Finv_{X_T^\delta}(u)}
{X_0^\delta}-\Gamma\,T\right)
}{
\sqrt{
\left(
\lambda^\intercal\rho\lambda-\frac{(\lambda^\intercal\rho\eta)^2}{\eta^\intercal\rho\eta}
\right)
T
}
}
\right).
\end{split}
\end{equation}
Inserting \eqref{eqn:sdf-gbm-const} and \eqref{eqn:bench-gbm-const}, we therefore have
\begin{equation}
\tilde{U}:=F_{\varsigma_T|U^\delta}(\varsigma_T|U^\delta)
= \Phi\left(\tfrac{1}{\sqrt{T}}\theta^\intercal W_T\right),
\qquad
\text{where}
\qquad
\theta:=\frac{-\lambda+\frac{\lambda^\intercal\rho\eta}{\eta^\intercal\rho\eta}\eta}
{
\sqrt{
\left(
\lambda^\intercal\rho\lambda-\frac{(\lambda^\intercal\rho\eta)^2}{\eta^\intercal\rho\eta}
\right)
}
}.
\end{equation}
Note, after some tedious computations, one can show that  $\theta^\intercal\rho\theta = 1$, hence  $\frac{1}{\sqrt{T}}\theta^\intercal W_T \stackrel{\mathbb{P}}{\sim} \mathcal N(0,1)$ and so $\tilde{U} \stackrel{\mathbb{P}}{\sim} U(0,1)$ as required. Moreover, the covariance $\mathbb{C}\left[\eta^\intercal W_T;\theta^\intercal W_T\right] = \eta^\intercal\rho\theta\;T=0$ which implies that  $\tilde{U}$ and $U^\delta$ are independent.

Next, from the definition of $V$,
\begin{subequations}
\begin{align}
\Q(V\le v) 
&= \Q\left( \Coninv\left(1 - \tU\, |\,  U^\delta\right) \le v \right)
= \Q\left( 1 - \tU\, \le \Con\left(v\, |\,  U^\delta\right) \right)
\\
&= \Q\left( \tU\, >  1 -\Con\left(v\, |\,  U^\delta\right) \right)
= 1 - \Q\left( \tU\, \le  1 -\Con\left(v\, |\,  U^\delta\right) \right).
\end{align}
\label{eqn:Q_V}
\end{subequations}
We thus require the $\Q$-distribution of $\tU$ and $U^\delta$ -- they are $\P$-independent and therefore $\Q$-independent. First,
\begin{subequations}
\begin{align}
\Q(\tU\le u) 
&= \Q\left(F_{\varsigma_T|U^\delta}(\varsigma_T|U^\delta)\le u\right)
= \Q\left(\Phi\left(\tfrac{1}{\sqrt{T}}\theta^\intercal W_T\right)\le u\right)
\\
&= \Q\left(\tfrac{1}{\sqrt{T}}\theta^\intercal W_T\le \Phi^{-1}(u)\right)
= \Q\left(\tfrac{1}{\sqrt{T}}\theta^\intercal (-\rho\,\lambda\,T+W_T^*)\le \Phi^{-1}(u)\right)
= \Phi\left( \Phi^{-1}(u)+\theta^\intercal\,\rho\,\lambda\,\sqrt{T}\right).
\end{align}%
\label{eqn:Q_Ut}%
\end{subequations}%
The last equality follows as $\theta^\intercal\rho\theta=1$, so that $\frac{1}{\sqrt{T}}\theta^\intercal W_T^*\stackrel{\Q}{\sim}\mathcal N(0,1)$. Second,
\begin{subequations}
\begin{align}
\Q(U^\delta\le u) &= \Q\left(F_{X_T^\delta}(X_T^\delta)\le u\right)
= \Q\left(
X_T^{\delta} \le X_0^\delta\,e^{\Gamma\,T+ \Psi\,\sqrt{T}\,\Phi^{-1}(u)}
\right)
\\
&= \Q\left(\frac{\eta^\intercal W_T}{\Psi\,\sqrt{T}} \le \Phi^{-1}(u)\right)
= \Q\left(\frac{\eta^\intercal (-\rho\lambda\,T+W_T^*)}{\Psi\,\sqrt{T}} \le \Phi^{-1}(u)\right)
= \Phi\left( \Phi^{-1}(u) + \vartheta^\intercal\rho\lambda\,\sqrt{T}\right).
\end{align}%
\label{eqn:Q_U}%
\end{subequations}%
Putting \eqref{eqn:Q_V},\eqref{eqn:Q_Ut}, and \eqref{eqn:Q_U} together, we arrive at \eqref{eqn:Q_V_GBM}.
\Halmos

\section{Model Parameters for Simulations}
\subsection{GBM market model}\label{app:GBM-param}

The GBM market models consist of two risky assets with 
\begin{equation}
    \mu = \begin{pmatrix}0.05 \\ 0.06 \end{pmatrix},
\quad \sigma =  \begin{pmatrix} 0.1 \\ 0.12 \end{pmatrix},
\quad
\rho =  \begin{pmatrix} 1 &  0.25 \\ 0.25 & 1 \end{pmatrix},
\quad
\delta_t = \begin{pmatrix} 0.25\\0.75 \end{pmatrix},
\quad
\text{and}
\quad
S_0 = \begin{pmatrix}1 \\ 2\end{pmatrix},
\end{equation}
where the drift and volatility are expressed on an annual basis. The interest rate is constant equal to $r = 0.01$ and the investment horizon is $T = 5$ years. The investor's initial wealth  is $X_0^\delta = 1$.

\subsection{SIR-CEV market model}\label{app:SIR-MM-param}
The SIR-CEV market model, see Example \ref{ex: SIR-MM}, used for simulations consists of two equities and a bond with parameters under the $\P$-measure
\begin{subequations}
\begin{equation}
    \mu = \begin{pmatrix}0.05 \\ 0.06 \end{pmatrix},
\quad 
\sigma =  \begin{pmatrix} 0.2 \\ 0.32 \end{pmatrix},
\quad
\beta = \begin{pmatrix} -0.2 \\ -0.3 \end{pmatrix},
\quad
\rho =  \begin{pmatrix} 1 &  0.25 & 0.2 \\0.25 & 1 &  0.3 \\ 0.2 & 0.3 & 1 \end{pmatrix},
\quad
\delta_t = \begin{pmatrix} 0.2\\0.6\\0.1 \end{pmatrix},
\quad
\text{and}
\quad
S_0 = \begin{pmatrix}1 \\ 2\end{pmatrix}\,.
\end{equation}
The interest rates satisfies $r_0 =0.02$ with 
\begin{align}
&\text{$\P$-parameters}
    \quad
    \kappa = 1, 
    \quad
    \theta = 0.02,
    \quad
    \sigma_r = 0.02,
    \quad \text{and}\\
&\text{$\Q$-parameters }
    \quad
    \kappa = 1, 
    \quad
    \theta = 0.025,
    \quad
    \sigma_r = 0.02\,.
\end{align}
where the drift and volatility are expressed on an annual basis. The investor's initial wealth  is $X_0^\delta = 1$ and the investment horizon of $T = 5$ years.
\end{subequations}

\section{Strategies Distinct from Benchmark}
\label{sec:distinct_strategies}

\new{
Here, we prove that there always exists a strategy that satisfies properties (i), (ii), and (iii) of Definition \ref{def:optimal-strategy} that is distinct from the benchmark.

Let $X_T^\delta \stackrel{\mathbb{P}}{\sim} F$ be the terminal wealth random variable of the benchmark strategy. First, consider the case when the copula $\Co$ is not the comonotonic copula. Then set $X_T^\pi : = F^{-1}(V)$ with $V$ given in Equation \eqref{eq:V}. Then, by Proposition \ref{prop:quantile-rep}, $(X_T^\pi, X_T^\delta)$ have copula $\Co$, the Wasserstein distance between $X_T^\pi$ and $X_T^\delta$ is equal to zero as they have the same distribution, i.e. $d_2[X_T^\pi, X_T^\delta] = 0$, and the budget constraint is satisfied since from Lemma \ref{lemma-optimal-budget} it holds that $\mathbb{E}[\varsigma_T\,X_T^\pi ] = \mathbb{E}[\varsigma_T\,F^{-1}(V) ] < \mathbb{E}[\varsigma_T\,X_T^\pi ]  = X_0^\delta$. We remark that $X_T^\pi$ is distinct from $X^\delta_T$ despite that they share the same distribution, as their copula is not the comonotonic one, and hence the corresponding replicating strategies differ.

Next, consider the case when the specified copula $\Co$ is the commonotonic copula. Then  set $X_T^\pi := X_T^\delta - \sqrt{\varepsilon}$, in which case $(X_T^\pi, X_T^\delta)$ are comonotonic ($X_T^\pi$ is an increasing function of $X_T^\delta)$, the Wasserstein distance is $d_2[X_T^\pi, X_T^\delta] = \E[\left(X_T^\pi - X_T^\delta)^2\right] = \varepsilon$, and the budget constraint is $\mathbb{E}[\varsigma_T \, X_T^\pi] = \mathbb{E}[\varsigma_T \, X_T^\delta] - \sqrt{\varepsilon} < X_0$.
As $X_T^\pi$ is a constant shift of $X_T^\delta$, the corresponding replicating strategies differ.

Therefore, there always exists a strategy $\pi \in \mathcal{A}$, that satisfies condition (i), (ii), and (iii) of Definition \ref{def:optimal-strategy}  and is not equal to the benchmark.

}

\end{APPENDICES}





\bibliographystyle{informs2014}
\bibliography{RefsBeatingBenchmark.bib}

\end{document}